\documentclass[aps,pra,superscriptaddress,floatfix,footinbib,twocolumn,10pt]{revtex4-1}
\pdfoutput=1
\usepackage{graphicx}
\usepackage{amsmath,amssymb,mathrsfs}
\usepackage{bbm}
\usepackage{hyperref}
\usepackage{gensymb}       
\usepackage{paralist}

\begin{document}


\title{Unveiling two types of local order in liquid water using machine learning}

%


%
\author{Adri\'an Soto}
\email{adrian.soto-cambres@stonybrook.edu}
\affiliation{Department of Physics and Astronomy, Stony Brook University, Stony Brook, NY 11794-3800}
\affiliation{Institute for Advanced Computational Science, Stony Brook University, Stony Brook, NY 11794}

\author{Deyu Lu}
\email{dlu@bnl.gov}
\affiliation{Center for Functional Nanomaterials, Brookhaven National Laboratory, Upton, NY 11793}

\author{Shinjae Yoo}
\email{syoo@bnl.gov}
\affiliation{Institute for Advanced Computational Science, Stony Brook University, Stony Brook, NY 11794}
\affiliation{Computational Science Initiative, Brookhaven National Laboratory, Upton, NY 11793}

\author{Mariv\'i Fern\'andez-Serra}
\email{maria.fernandez-serra@stonybrook.edu}
\affiliation{Department of Physics and Astronomy, Stony Brook University, Stony Brook, NY 11794-3800}
\affiliation{Institute for Advanced Computational Science, Stony Brook University, Stony Brook, NY 11794}

\preprint{} 


\begin{abstract}
Machine learning methods are being explored in many areas of science, with the aim of finding solution
to problems that evade traditional scientific approaches due to their complexity. 
In general, an order parameter capable of identifying two different phases of matter separated by
a corresponding  phase transition is constructed based on symmetry arguments. This
 parameter measures the degree of order as the phase transition proceeds.
However, when the two distinct phases are highly disordered it is not trivial to identify broken symmetries
with which to find an order parameter. This poses an excellent problem to be addressed
using machine learning procedures. Room temperature liquid water is hypothesized to be a supercritical liquid,
with fluctuations of two different molecular orders associated to two parent liquid phases, one with high density and another one with low density. The validity of this hypothesis is linked to the existence of an order parameter
capable of identifying the two distinct liquid phases and their fluctuations.
In this work we show how two different machine learning procedures are capable of recognizing
local order in liquid water. We argue that when in order to learn relevant features from this complexity, an initial, physically motivated preparation of the available data
is as important as the quality of the data set, and that machine learning can become a successful analysis
tool only when coupled to high level physical information.

\end{abstract}

\maketitle

\section{Introduction}

Perhaps the most important open debate about the physics of water is the existence of a liquid-liquid phase transition, hypothesized 25 years ago \cite{Stanley1992}. The existence of this phase transition with a critical point at supercooled temperatures and high pressures provides a mechanism to explain some of the anomalous macroscopic properties of water, such as the decrease of the heat capacity and the isothermal compressibility of water at ambient conditions upon heating \cite{Millero1969} or the anomalous behavior of the density with temperature, reaching a maximum at $4 \degree$C and showing a negative coefficient of thermal expansion below this temperature into the supercooled regime \cite{Speedy1974}. In this picture, water molecules above the critical temperature fluctuate between a low density, enthalpy-favored environment with a highly tetrahedral coordination and a high density, entropy-favored environment with highly distorted structures (we adopt the standard nomenclature low density/high density (LD/HD) for the former/latter respectively). The ratio of the HD to LD populations increases with temperature, yielding a normal liquid behavior at high temperatures but an anomalous behavior from the vicinity of ambient conditions down in temperature into the supercooled region \cite{Nilsson2015}, where the populations of the two types of molecules becomes comparable. 

Since it was proposed, this hypothesis has motivated intensive experimental and computational investigations on the existence of a dual microscopic nature of water. X-ray spectroscopy experiments have shown that pre- and post-edge peaks, which characterize distorted H-bonds and strong H-bonds respectively, have a distribution whose temperature dependence agrees with the above hypothesis \cite{Huang2009}. 
Recently, novel X-ray experiments observed a continuous transition between HD liquid and LD liquid at ambient pressure and temperatures between $110$ K and $130$ K, agreeing with the hypothesis of a first order liquid-liquid phase transition at elevated pressures \cite{Perakis2017}.
Molecular dynamics (MD) simulation studies of liquid water also suggest this trend \cite{Giovambattista2012, Liu2012, Russo2014}.
Furthermore, recent studies  \cite{Wikfeldt2011, Santra2015} show that the inherent structures (IS) of MD trajectories of water, which are obtained by relaxing the finite temperature structures to their corresponding potential energy minima, display a bimodal distribution of the Local Structure Index (LSI) \cite{Sasai1996}. The sizes of the modes of the distribution vary with temperature and pressure of the finite temperature (unrelaxed) simulation, supporting the hypothesis of the existence of two molecular environments even at temperatures well above the predicted critical temperature. 

This prevaling view, however, has been heavily contested. A lack of evidence of heterogeneity in simulations \cite{Moore2009, Clark2010, Limmer2011, Limmer2013, Szekely2016} and experiments \cite{Soper2010, Clark2010} has been pointed out, attributing the seemingly biphasic character to noncritical thermal fluctuations.
Over the past couple of decades the community has been putting great effort into developing local order parameters for water. These capture different properties such as packing, tetrahedral arrangement and angular distribution of the first few neighboring molecules \cite{Sasai1996, Chau1998, Bako2013, Laage2015}. While the LSI is bimodal at the IS, none of them has successfully shown the expected bimodality when evaluated at the finite temperature structures near ambient conditions.

Therefore the extraction of local order parameters in liquid water presents an ideal problem to treat with machine learning due to its capability of unveiling information from complex data. Recent studies have successfully utilized machine learning in condensed matter physics to study structure and order: in Ref. \cite{Carrasquilla2017} a convolutional neural network was trained to successfully classify Monte Carlo-drawn configurations of two-dimensional Ising models above and below the critical point, and in Ref. \cite{Ceriotti2014} unsupervised learning was used to construct an agnostic, data-driven definition of a hydrogen bond in ice, water and ammonia based only on structural information. These developments prove that machine learning can be used for phase identification as well as for finding subtleties in the geometrical structure of complex three-dimensional systems. 

Inspired by these studies, in this paper we propose a method for finding local order parameters with machine learning which, if successful, will provide a more flexible means to classify local structures of complex materials in bulk but also in more challenging scenarios such as near interfaces or at extreme thermodynamic conditions. We then turn our attention to liquid water, where we employ this machine learning methodology to study whether the IS are truly revealing two types of molecular arrangements based on local geometrical information. Finally we discuss the effects of statistical fluctuations.

\section{Methods}

\subsection{Formalism}

\begin{figure}
	\includegraphics[width=1.0\linewidth]{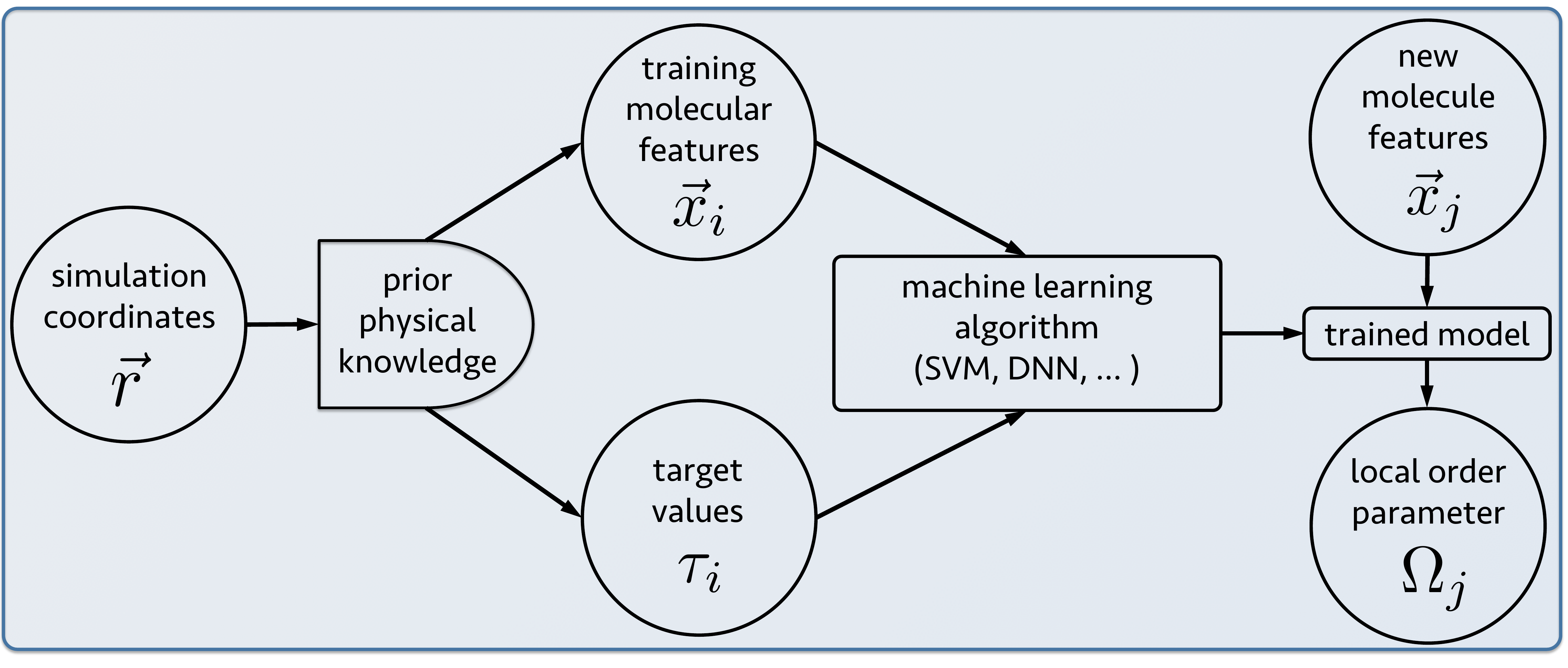}
    	\caption{Machine learning local order parameter workflow. Feature vectors describing the molecular environment and target values describing distinct molecular classes need to be provided by physical intuition. The machine learning algorithm learns from these data, allowing to characterize a new molecular environment.} \label{fig:MLchart}
\end{figure}

Let $\vec r_t = \left( \vec r_{1,t}, \vec r_{2,t}, \cdots , \vec r_{n,t} \right)$ be the molecular (or atomic) coordinates at simulation at step $t$ \footnote{$t$ can be a time step in MD or a trial index if Monte Carlo methods are used to sample configurational space.}. In the following, we drop the step index $t$ and write it explicitly when necessary. 
Let us define a \textit{feature vector} $\vec x_i = \vec x_i \left( \vec r \right)$ of dimension $D$ whose components contain information about the geometry of molecule $i$ and its environment. In addition, a target value that characterizes the distinct molecular environments needs to be provided in order to perform supervised machine learning. Both the feature vector and the target should be specified from educated physical intuition. These two elements can then be combined to train a machine learning algorithm and machine learn a local order parameter $O$. A schematic representation of this procedure is shown in Fig. \ref{fig:MLchart}.

Let us now focus now on the procedure that will be employed in the rest of this work to obtain the target values. Assume that we know a local order parameter, $\tilde O_i \left( \vec r \right)$, function of the molecular coordinates only, whose probability distribution
$P ( \tilde O ( \vec{r} ) )$ is bimodal, thereby allowing to interpret each of the modes of the distribution as a distinct class of local order. Based on this, one could attempt to machine learn a function $ O ( \vec x_i )$ that approximates the order parameter $\tilde O_i  \left( \vec r \right)$. Notice that $O$ is an explicit function of the local features $\vec x$ only and implicitly of the atomic coordinates $\vec r$
\footnote{Note that the original order parameter can be trivially recast in this new formalism by setting $\vec x ( \vec r )= \vec r$ and $O = \tilde O_i$}.
Alternatively, instead of a continuous order parameter one could learn a binary output function, with one output value for each of the two peaks of the probability distribution $P ( O )$. With an appropriate feature vector $\vec x$, a machine learning algorithm can find the desired approximate order parameter $O$ provided that the local features contain the necessary information.

In order to bring this formalism to practice the following aspects need to be properly regarded: 
\begin{inparaenum}[(i)]
	\item appropriate sampling of the thermodynamic ensemble,
	\item feature set choice that contains enough information to describe the binary character of the molecular environment, and 
	\item a suitable machine learning model, training scheme and hyperparameter optimization.
\end{inparaenum}  
In the following we describe how we approach each of those items.

\subsection{Ensemble sampling with molecular dynamics} 
We performed a series of classical MD simulations in the canonical (NVT) ensemble at ambient density ($1.0$ kg/l) and at seven different temperatures. We employed the TTM3-F force field \cite{TTM3Fpaper}, which is a flexible and polarizable water model fitted to MP2/aug-cc-pVDZ calculations of water clusters, yielding very good agreement with experiment in the structure and the vibrational spectra of liquid water at $T = 298\text{ K}$. Our simulations were performed using a cubic periodic cell of length $L=15.64 \text{ Ang}$ containing 128 molecules. This cell allows us to study correlations of up to $\sim 7.8 \text{ Ang}$, sufficient for our study of local order in the $\lesssim 6.0\text{ Ang}$ range. We control the temperature by means of a Nos\'e-Hoover thermostat \cite{NH-Nose, NH-Hoover}. 

All of our MD productive runs are $50$ ps long with a Verlet integration step of $0.5$ fs following $50$ ps of equilibration. The simulations were performed with our own MD code \cite{DanMD}.

From each of those simulations $2000$ snapshots were relaxed using the conjugate gradient method \cite{Stiefel1952}, converging the maximum atomic force below $5$ meV/Ang. 

\subsection{Learning order from inherent structures}
The partition function of a classical liquid can be written (up to a constant kinetic prefactor) as \cite{Stillinger1982}
\begin{equation}\label{eq:SWpartitionfunction}
	Z_N \sim \sum_\alpha \frac{1}{\sigma_\alpha} e^{-\beta V_\alpha} \int_{R_\alpha} d^{3N} \vec r \: e^{-\beta \Delta V_\alpha \left( \vec r \right)},
\end{equation}
where the sum runs over all the local minima of the potential energy surface, $\alpha$. $V_\alpha$ is the potential energy at minimum $\alpha$, and $\Delta_\alpha V \left( \vec r \right) = V \left( \vec r \right) - V_\alpha $ is the energy difference above the local minimum $\alpha$ to which the system configuration with coordinates $\vec r$ would relax. Each integral is performed over the coordinate space subvolume $R_\alpha$ which is the coordinate relaxation basin of attraction of the points $\vec r_\alpha$. $\beta = 1/k_BT$ is the usual inverse temperature and $\sigma_\alpha$ is a translational symmetry number (in our case, since we have periodic boundary conditions, $\sum_\alpha \sigma_\sigma = N_{\text{particles}}$). The coordinates at the local minima of the potential energy are the IS and they play a crucial role in our analysis as we shall see shortly. 

The LSI is defined as
\begin{equation}\label{eq:LSI}
	I = \frac{1}{N_s}\sum_{j=i}^{N_s} \left( \delta_{j+1,j} - \left\langle \delta \right\rangle \right)^2
\end{equation}
where the sum is carried out over the $N_s$ pairs of molecules within a cutoff shell of radius $r_c = 3.7$ Ang. $\delta_{j+1,j}=r_{j+1}-r_j$ is the difference in radii of the shells occupied by molecules $j$ and $j+1$, and $\left\langle \delta \right\rangle = \frac{1}{N_s}\sum_{j=i}^{N_s} \delta_{j+1,j}$ is the average radii difference. This quantity is designed to measure the degree to which first and second H-bond coordination shells in water are well-defined, with low LSI values for distorted coordination shells and high LSI values for very regular coordination shells.

\begin{figure}
	\includegraphics[width=0.9\linewidth]{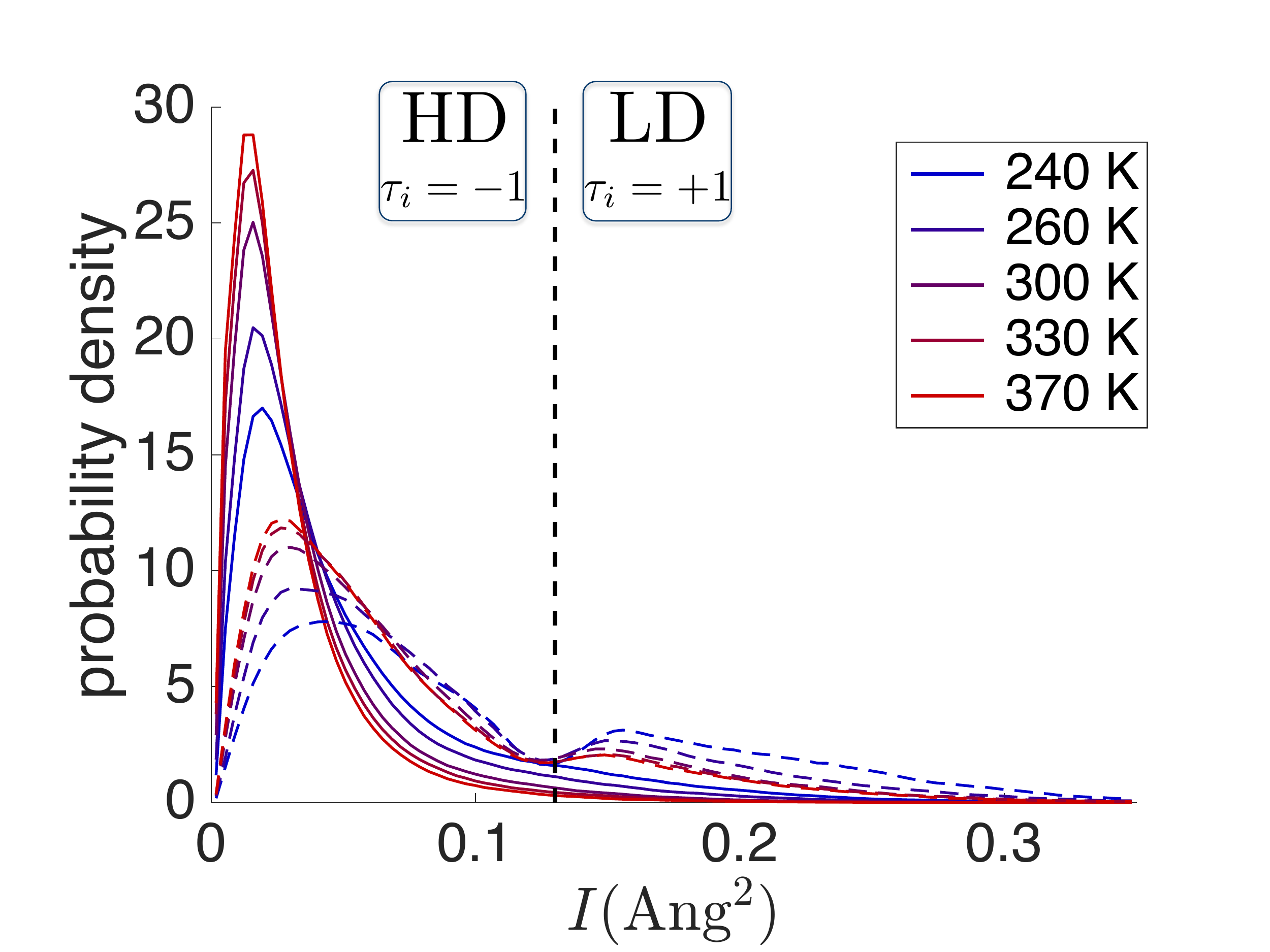}
	\vspace{-0.4cm}
    \caption{Normalized histograms of the LSI evaluated at finite temperature (solid lines) and at the inherent structure (dashed lines). The simulation temperatures range from $240$ K (blue) to $370$ K (red). A threshold value of $I_0=0.13\text{ Ang}^2$ separating the HD from the LD target values is shown with a vertical dashed line.} \label{fig:hist_LSI_withtags}
\end{figure}

\begin{figure*}[]
	\centering
	\includegraphics[width=0.240\linewidth]{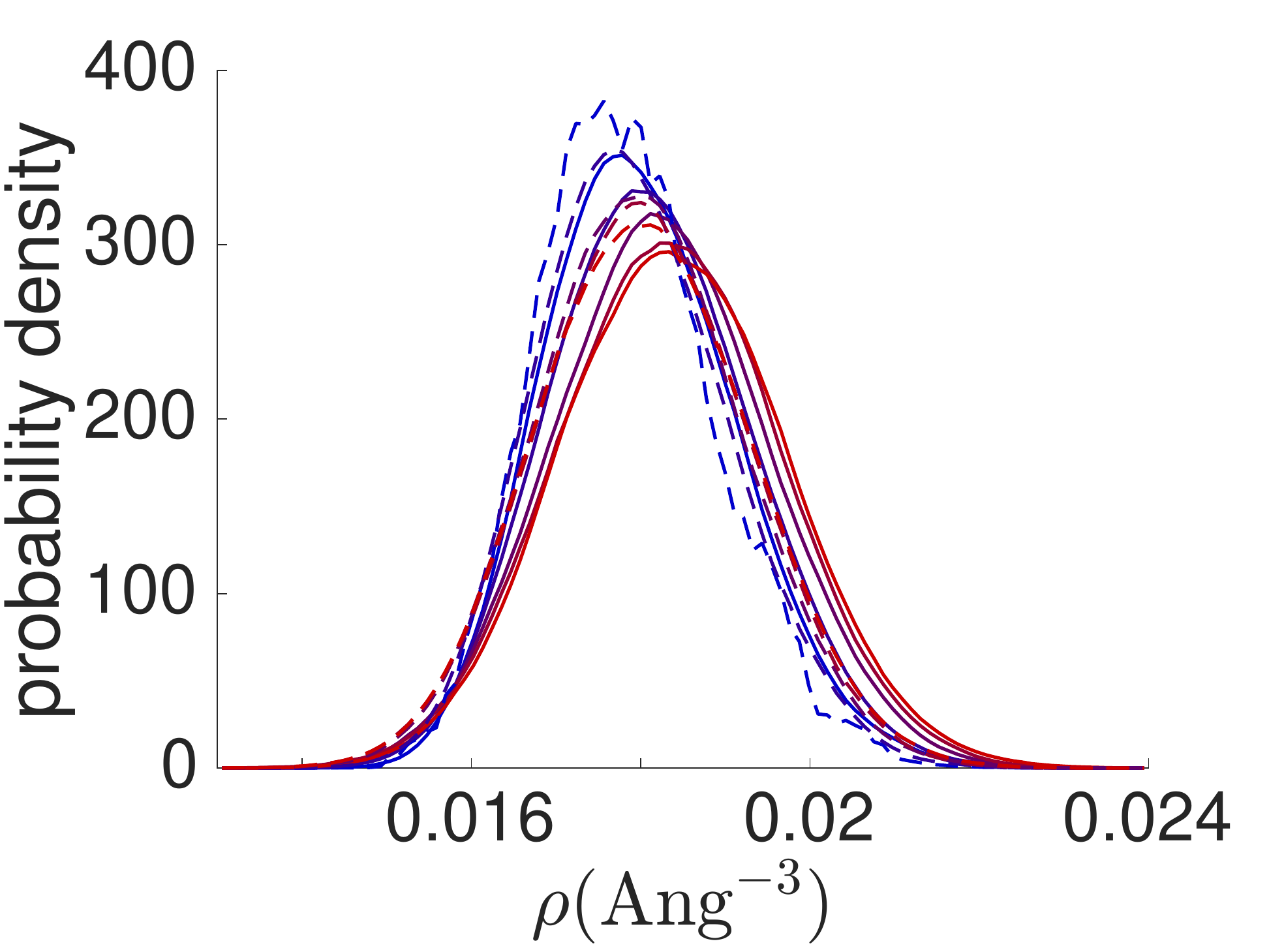} 
	\includegraphics[width=0.240\linewidth]{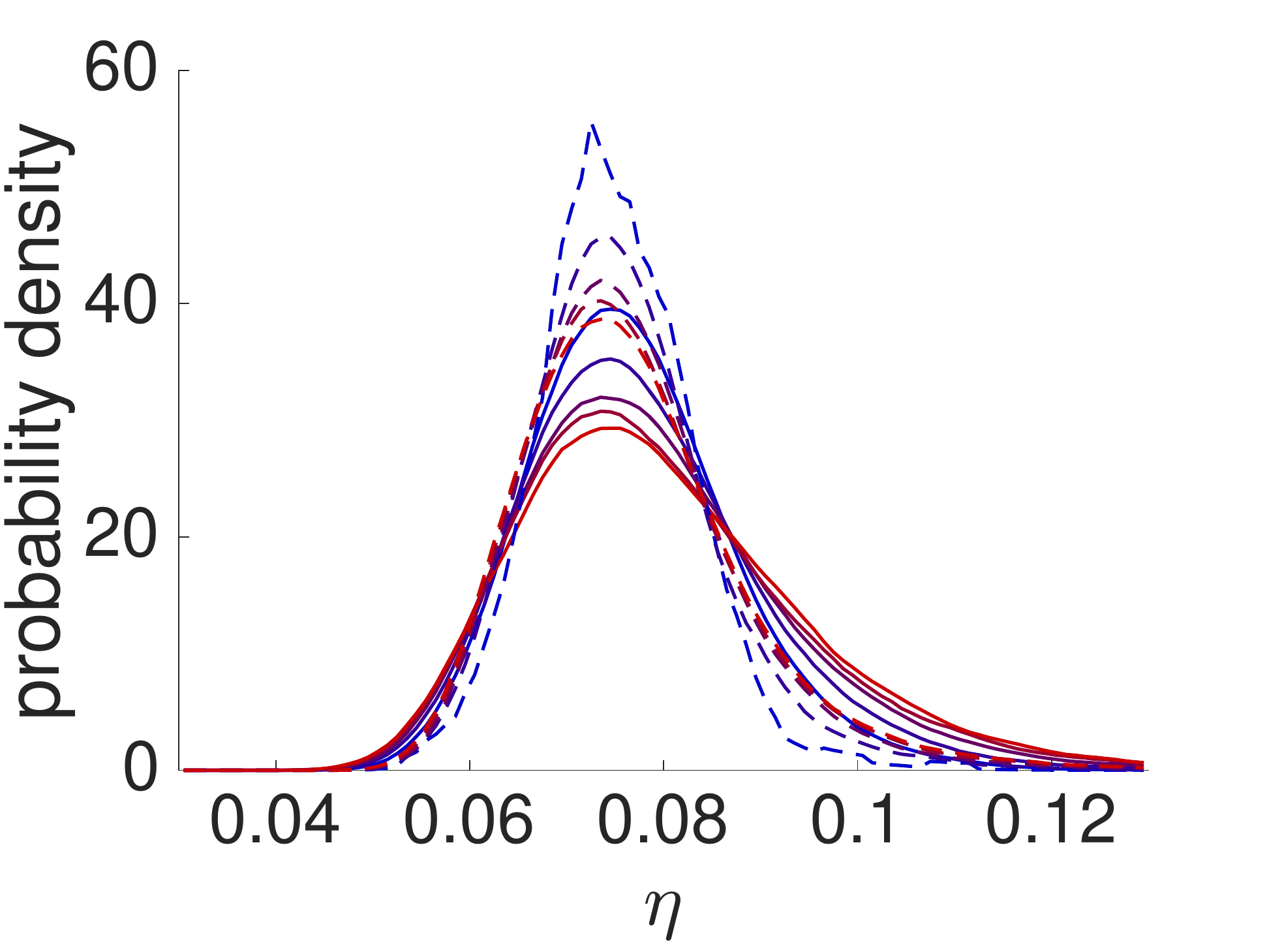} 
	\includegraphics[width=0.240\linewidth]{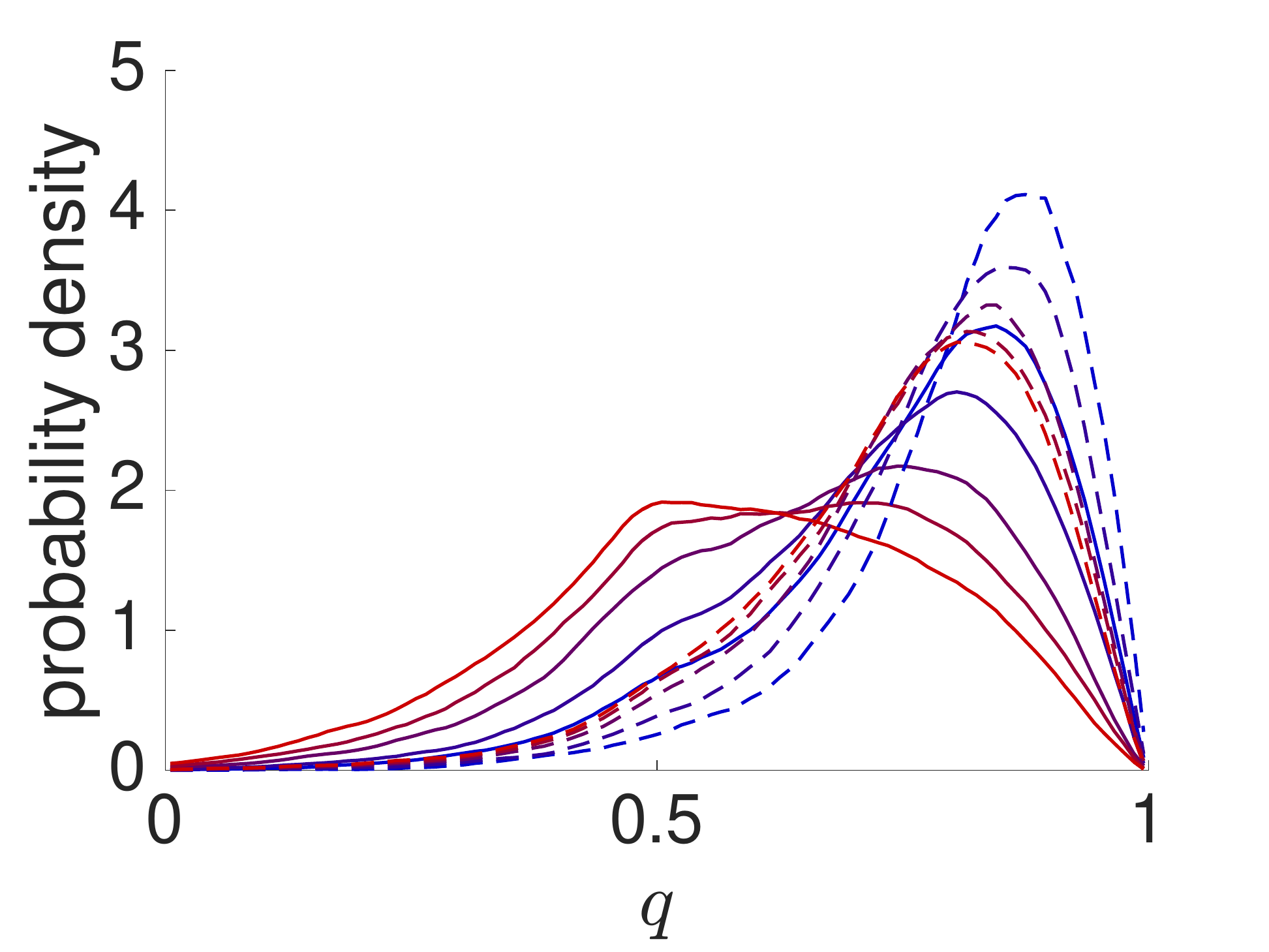} 
	\includegraphics[width=0.240\linewidth]{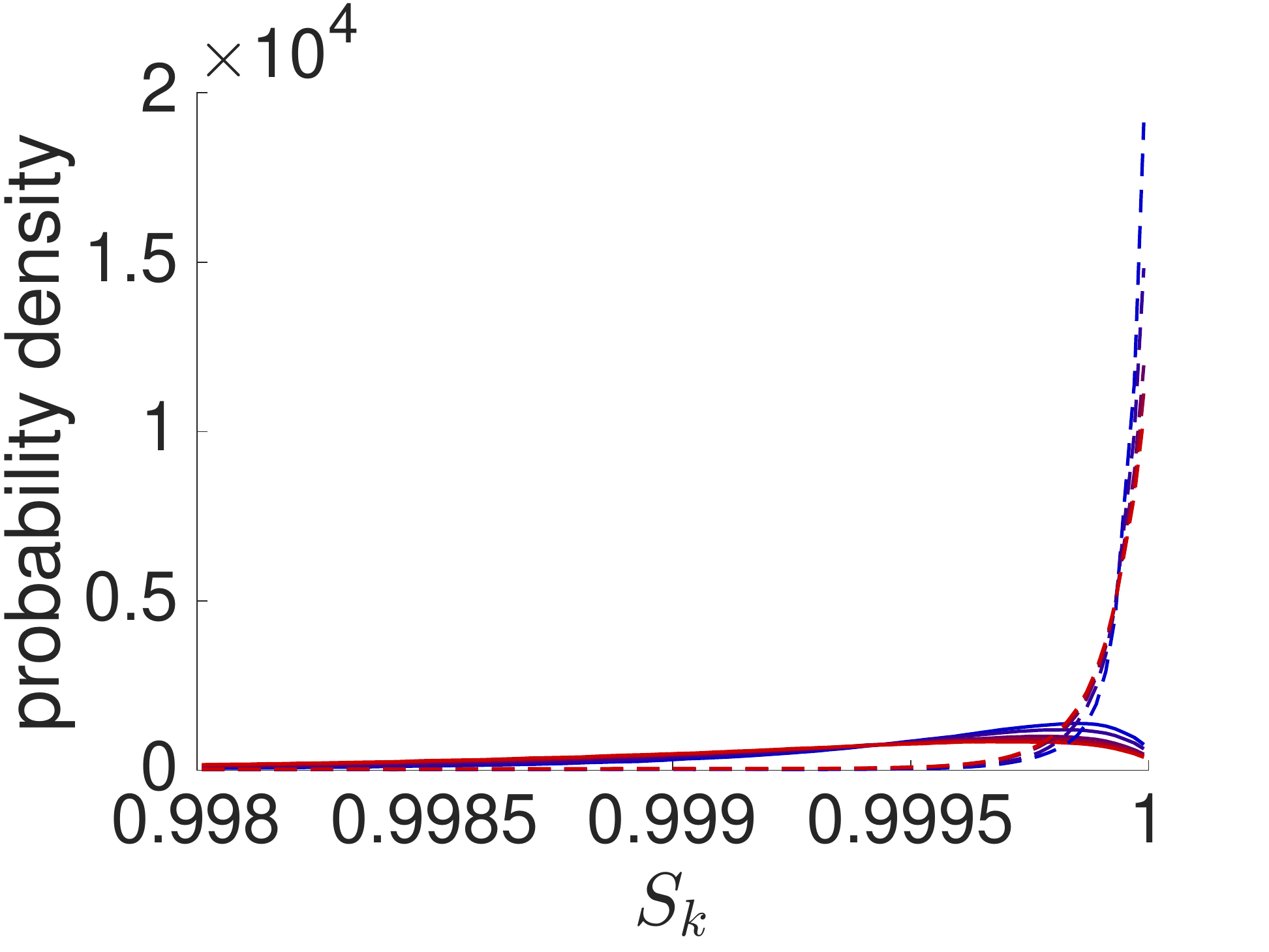} 
	\medskip
	\vspace{-0.4cm}
	\caption{From left to right: normalized histograms of Voronoi cell inverse volume, $\rho$, Voronoi cell asphericity, $\eta$, orientational tetrahedral order parameter, $q$, and translational tetrahedral order parameter, $S_k$, evaluated at finite temperature (solid lines) and at the inherent structure (dashed lines). The simulation temperatures range from $240$ K (blue) to $370$ K (red). No evident bimodality is observed in these quantities at the IS or at finite temperature.} 
	\label{fig:HistPanel}

\end{figure*}

We calculate the \textit{inherent LSI}, the LSI on the relaxed atomic coordinates, at several different simulation temperatures. While the finite temperature (FT) LSI distribution is unimodal, the inherent LSI distribution in Fig. \ref{fig:hist_LSI_withtags} shows a bimodal character. As it can be seen in Fig. \ref{fig:HistPanel} this is not the case for other suggested order parameters. The inherent LSI distribution shows a decreasing LD population with increasing temperature, and it has been argued that each mode characterizes a distinct class of molecular environment suggesting that the inherent LSI is a useful local order parameter\cite{Wikfeldt2011, Santra2015, Nilsson2015}. Following these studies we propose to use inherent structures of water to characterize their corresponding finite temperature structures. Since the location of the maximum separating the two peaks of $P(I)$ does not change substantially with the simulation temperature, we can apply a unique threshold value $I_0$ to separate the two modes, each representing their HD or LD character respectively (dashed lines in Fig. \ref{fig:hist_LSI_withtags}). This provides a way to assign binary target values of $-1$ (HD) or $+1$ (LD) to each data point, enabling us to perform supervised machine learning. In this work we choose the threshold value to be at the minimum between the two modes of the distribution, $I_0=0.13 \text{ Ang}^2$, in agreement with similar studies with TIP4P/Ice force field and slightly below values $I_0=0.14-0.16 \text{ Ang}^2$ obtained from AIMD \cite{Santra2015}.

\subsection{Molecular features and target values}\label{ss:Features}
\noindent From the atomic coordinates of the MD trajectories we compute the local features of all the molecules in our sample and aggregate them into feature vectors $\vec x_i$. In this work we construct our feature vectors with the following local geometrical quantities: 
\begin{inparaenum}[(I)]
	\item inverse volume $\rho$, asphericity $\eta$, number of sides $s$, edges $e$ and vertices $v$ of the Voronoi cell of each molecule, determined by the position of its O atom \cite{Voro++Paper},
	\item orientational and translational tetrahedral order parameters $q$ and $S_k$ \cite{Chau1998},
	\item local structure index $I$ \cite{Sasai1996},
	\item 17 intermolecular O-O distances,
	\item 5 O-O-O angles,
	\item  6 O-H--O angles,
	\item \label{Item:numH-bonds} number of donated and accepted H-bonds and
	\item \label{Item:numH-bondLoops} number of H-bond loops of lengths 3-12 that the $i$ participates in \cite{Bako2013}.
\end{inparaenum}
In order to evaluate \ref{Item:numH-bonds} and \ref{Item:numH-bondLoops} we choose a standard geometrical definition of the H-bond in which two molecules are hydrogen bonded if $r_{\textit{OO}} < 3.5$ Ang and $\widehat{ O_{\text{a}} O_{\text{d}} H_{\text{d}} } < 30^{\circ}$ \cite{Kumar2007, Corsetti2013}. The first condition ensures that the two molecules are within the first coordination shell of the pair distribution function and the second that the molecules are properly aligned to form an H-bond. 

In order to identify which of these features are most relevant to a successful description of the molecular environment, we select different combinations of them, isolating local order parameters and Voronoi cell quantities from the local H-bond network topology and from intermolecular distances and angles. 
The feature set choices utilized in this study are listed in Table \ref{table:FeatureChoices}. 
\begin{table}[t]
\centering
\begin{tabular}{c|c|c|c|c|c|}
\cline{2-6}
                              & A & B & C & D      & E \\ \hline
\multicolumn{1}{|c|}{I}       & y & y & y & n      & n                      \\ \hline
\multicolumn{1}{|c|}{II}      & y & y & y & n      & n                      \\ \hline
\multicolumn{1}{|c|}{III}     & y/n & y/n & y/n & n      & n                      \\ \hline
\multicolumn{1}{|c|}{IV}      & y & n & n & 6 n.n. & y                      \\ \hline
\multicolumn{1}{|c|}{V \& VI} & y & n & n & n      & y                      \\ \hline
\multicolumn{1}{|c|}{VII \& VIII}     & y & n & y & n      & n                      \\ \hline
\end{tabular}
\caption{Description of different local feature choices, A-E. The features are grouped by category as follows:
	(I) molecular Voronoi cell parameters,
	(II) tetrahedral order parameters,
	(III) LSI,
	(IV) O-O distances,
	(V) O-O-O angles,
	(VI) O-H--O angles,
	(VII) number of H-bonds and
	(VIII) number of H-bond loops 
(see \ref{ss:Features} for further details). The entries with value ``y/n" signify that that feature was used for finite temperature structure classification but not used for classification of inherent structures. 
Entries with a numeric value followed by ``n.n." denote up to how many nearest neighbors are included. If ``y" is specified, all of the features in that category are included.
}
\label{table:FeatureChoices}
\end{table}

\subsection{Order parameters from supervised machine learning} \label{ss:MLOFs}

\begin{figure}[b]
		\centering
		\includegraphics[width=0.9\linewidth]{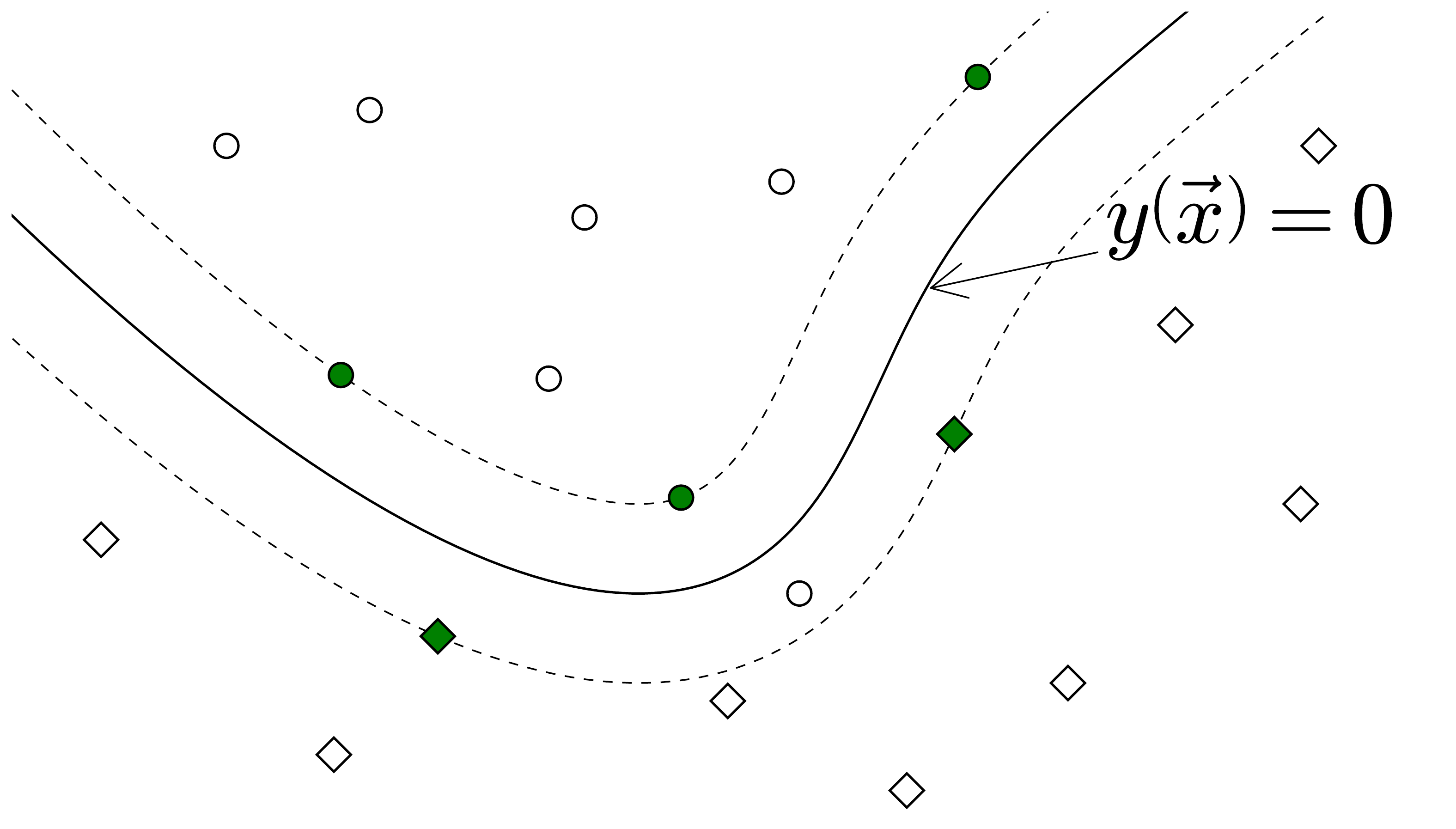}
		\caption{2d illustration of a SVM with a nonlinear kernel. It separates the two classes of data points (circles and diamonds) by finding the optimal decision hypersurface $y (\vec x ) = 0$. The dashed lines parallel to the decision surface represent the margins, on which the support vectors lie (filled circles and diamonds). When a point is misclassified, as the circle below the surface, it incurs a penalty in the cost function.}\label{fig:SVMdiagram}
\end{figure}

Once the feature vectors $\vec x_i$ and their corresponding target values $\tau_i$ are generated, we machine learn a local order parameter $O \left( \vec x \right)$. We use two different machine learning algorithms for this task: a support vector machine (SVM) and a deep neural network (DNN). 
SVMs \cite{NL-SVM} are a popular class of classification methods that operate by finding an optimal hypersurface in the space of features $\vec x$ that acts as boundary between two classes of data points (see Fig. \ref{fig:SVMdiagram}) \footnote{The data points that define the decision boundary are called \textit{support vectors} and the number of them is determined by the data and the algorithm. In the case of perfectly linear separation two support vectors are sufficient to generate the hyperplane (as they define the hyperplane normal), but in a more complicated scenario the number of support vectors will be much larger, often a significant fraction of the data set.}.
There are two hyperparameters in the SVM that require user tuning: the box parameter $C$, which sets the scale of the penalty of misclassified points during the training process (see Appendix \ref{app:SVM}), and the radial basis function width, $\sigma$, which determines the sharpness of the decision boundary in the neighborhood of the support vectors (see Appendix \ref{app:RBF}). It is customary to train and test the SVM over a grid of $\left( C, \sigma \right)$ values in order to choose the one that maximizes the classification accuracy
\begin{equation}\label{eq:ClassificationAccuracy}
	\mathcal{A} = 1 - M / N_p,
\end{equation}
where $M$ is the number of misclassified points and $N_p$ is the total number of data points. 

In a similar fashion to the SVM, supervised learning can be performed by means of deep neural networks (DNN). These machine learning algorithms have revolutionized the applications of machine learning in the recent years in fields as diverse as image recognition, natural language processing, and many others. They can learn highly nonlinear properties of the data with a relatively low algorithmic complexity. 
There is no standard or systematic procedure to choose a DNN architecture.
In general, a DNN with more layers has increased ability to perform complex nonlinear transformations of the features. On the other hand, a DNN with more neurons has higher number of degrees of freedom to perform these nonlinear operations.
While conceptually simpler, one restriction of DNNs over SVMs is that they tend to need a larger number of data points to be trained successfully. Therefore we will need to be cautious about our choices of network size and topology. We refer the reader to Appendix \ref{app:DNN} for more details on DNNs.
\begin{figure}
		\centering
		\includegraphics[width=0.9\linewidth]{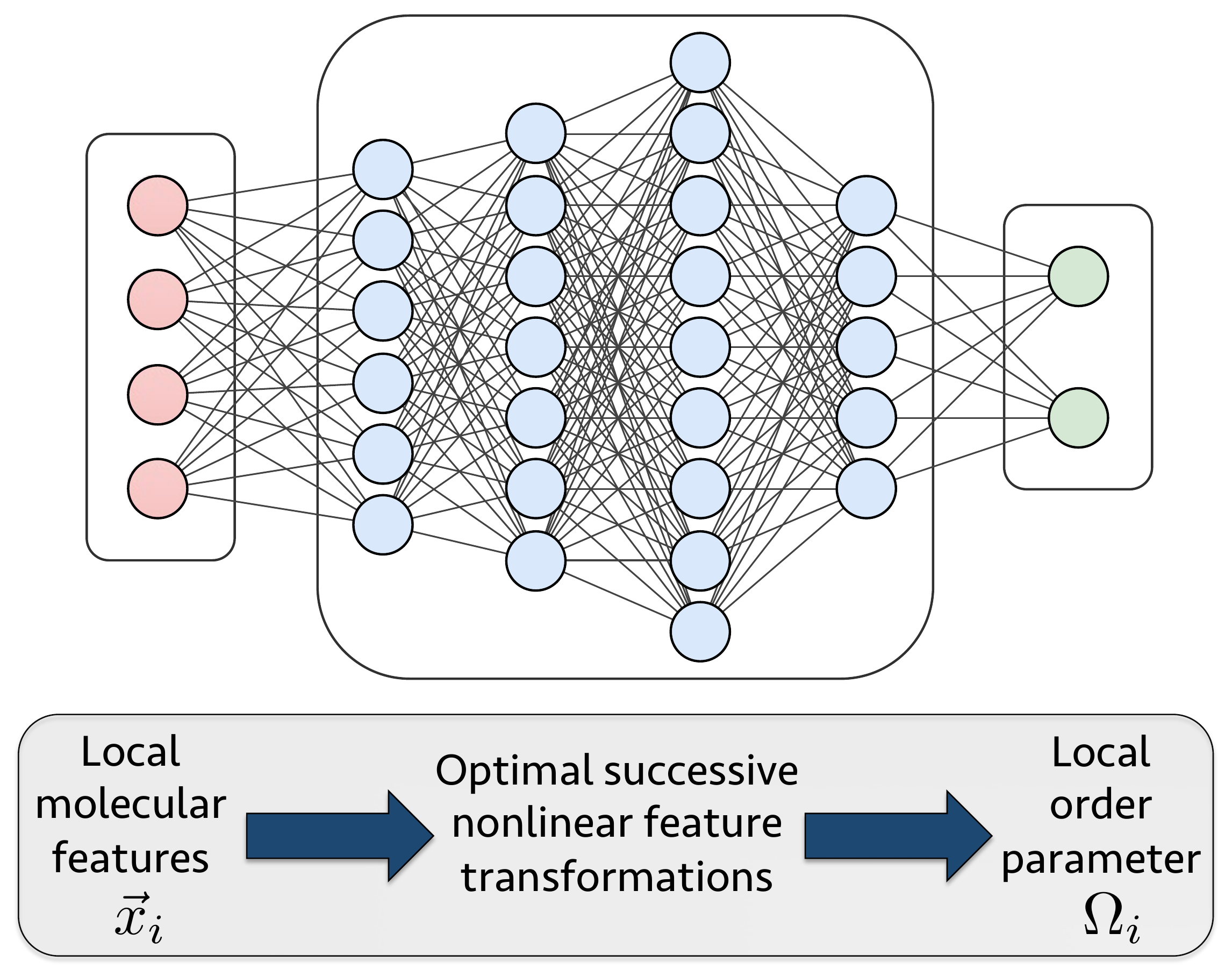}
		\caption{Schematic representation of our feed-forward deep neural network model for supervised learning local order parameters. The input nodes take feature vectors $\vec x_i$, which get computed forward into the network, connecting with all other features via nonlinear functions. The output layer contains as many nodes as classes, outputting the predicted class.}\label{fig:NNdiagram}
\end{figure}

We developed the codes to perform the SVM analyses with the MATLAB software package and the DNN were implemented with the open source Tensorflow package \cite{Tensorflow}. Example codes can be found online \cite{LocalOP-repo}. In Appendices \ref{app:SVM} and \ref{app:DNN} we provide more details about these methods for the interested reader.

\section{Results}

\subsection{Characterization of inherent structures}\label{ss:IS}

We begin by machine learning a local order parameter for inherent structures of liquid water. Note that in this case there is no need of using machine learning as we could directly calculate the inherent LSI from Eq. (\ref{eq:LSI}) and apply the threshold manually as discussed above. However, this will serve us as a proof of concept as well as a test case to assess the effectiveness of our procedure depending on the features we include in $\vec x$.
The data set is comprised of feature vectors from all our simulation temperatures. We begin from $N_{\text p}^{\text{init}}=1792000$ points, $256000$ from each temperature. It can be seen by the size of each mode  of the histogram in Fig. \ref{fig:hist_LSI_withtags} that our initial data set is unbalanced, meaning that $N_p(\tau = -1) \neq N_p(\tau = +1)$.  In order to prevent artificially biasing the classifier toward the most populated class (HD in our case), we balance the data set by randomly discarding a large number of HD points. 
From the balanced set, for SVM classification we randomly sampled $20000$ molecules from the remaining data for model training, $20000$ more for hyperparameter validation and another $20000$ for classification testing. 
In the case of the DNN, the computationally efficient batch gradient descent method allows to perform training with a significantly larger number of training points: $30000$ points were sampled for testing and all the remaining points were used for training. After multiple trials with several DNN number of layers and layer sizes, we chose a network with 4 hidden layers containing 80, 100, 200 and 75 neurons respectively (Fig. \ref{fig:NNdiagram}). 
%
\begin{table}[]
\centering
\begin{tabular}{|c|c|c|c|c|c|}
\hline
feature set choices & A     & B     & C     & D     & E     \\ \hline
$\mathcal{A}_\text{ SVM}$ & 0.96 & 0.74 & 0.78 & 0.99 & 0.97 \\ \hline
$\mathcal{A}_\text{ DNN}$ & 0.97 & 0.82 & 0.82 & 0.99 & 0.98 \\ \hline
\end{tabular}
\caption{Classification accuracies of a random mixture of inherent structures from initial simulation temperatures ranging between $240$ K and $370$ K obtained with a Support Vector Machine and with a Deep Neural Network for several feature choices.}
\label{table:AccuracyIS}
\end{table}

In Table \ref{table:AccuracyIS} we show the classification accuracies of classification of inherent structures of water as obtained with SVM and DNN. 
We first observe that both ML models produce similar errors across feature set choices A-E. The classification accuracy is $99\%$ for features D, showing that this method is able to identify very accurately to which of the HD and LD modes of the LSI distribution any given inherent structure belongs to if the only features in $\vec x$ are the first few intermolecular distances, which are the quantities that enter the LSI formula Eq. (\ref{eq:LSI}). 
When additional intermolecular distances and angles are included in $\vec x$, as in case E, the accuracy decreases slightly, but both models still provide an excellent classification. If the Voronoi cell information, tetrahedral order parameters and H-bond network structure features are included (case A), the accuracy decreases by about $1\%$. This is due to the noise that additional features introduce, which has a larger negative effect than the additional information they may provide relative to features C. Both D and E cases show that the introduction of additional features with irrelevant information can decrease the classification accuracy. In the case where intermolecular distances and angles are not included in $\vec x$, as in B and C, the accuracy decreases to about $75\% - 80\%$, showing that the tetrahedral order parameters, the single-molecule Voronoi cell parameters and the local H-bond topology features together do not contain enough information to identify reliably which of the two modes of the inherent LSI distribution a given molecule belongs to.

\subsection{Characterization of finite temperature structures}\label{ss:FT}
We now turn to the more challenging case of finite temperature (FT) structures. 
\begin{table}[]
\centering
\begin{tabular}{|c|c|c|c|c|c|}
\hline
feature set choice                 & A      & B      & C     & D     & E     \\ \hline
$\mathcal{A}_\text{ SVM}$ & 0.69 & 0.65 & 0.67 & 0.64 & 0.67 \\ \hline
$\mathcal{A}_\text{ DNN}$ & 0.74 & 0.65 & 0.63 & 0.64 & 0.69 \\ \hline
\end{tabular}
\caption{Classification accuracies of a random mixture of finite temperature structures from initial simulation temperatures ranging between $240$ K and $370$ K obtained with a Support Vector Machine and with a Deep Neural Network for several feature choices.} \label{table:AccuracyFT}
\end{table}
We use the same procedure as for the IS, with the only difference that the feature vectors $\vec x$ are constructed from FT structures. 
In Table \ref{table:AccuracyFT} we observe that this ML classification method yields classification accuracies below $75 \%$ for all feature set choices and for both ML models. The best results are obtained for A, where all the features are included. It is worth noting that, as opposed to the IS case, here the algorithm benefits from the information provided from additional features, yielding a higher accuracy for larger feature spaces. Very importantly, all feature set choices yield a better classification than that provided solely by the finite temperature LSI, which we find to give an accuracy of $0.60$. This proves that, despite the low overall accuracy, the ML algorithm is benefiting from the information provided in the other features to improve the classification accuracy relative to the FT LSI.

It is clear that this procedure struggles to accurately map FT local features to their corresponding HD/LD inherent LSI state. Formally, since inherent structures are derived from FT structures by a relaxation procedure $\mathcal R: \vec r  \rightarrow \vec r ^0 $, there should exist a one-to-one mapping from finite temperature structures to inherent structures, and therefore also to the inherent LSI. 
From Eq. (\ref{eq:SWpartitionfunction}), it is clear that as the system temperature increases and the structures start to oscillate around the inherent minima, the potential energy also increases, causing the partition function (and hence the free energy) to depart from that of the inherent structures. Therefore it is expected that 
the classification accuracy of our ML procedure would decrease with the system temperature.
\begin{figure}[]
	\includegraphics[width=0.9\linewidth]{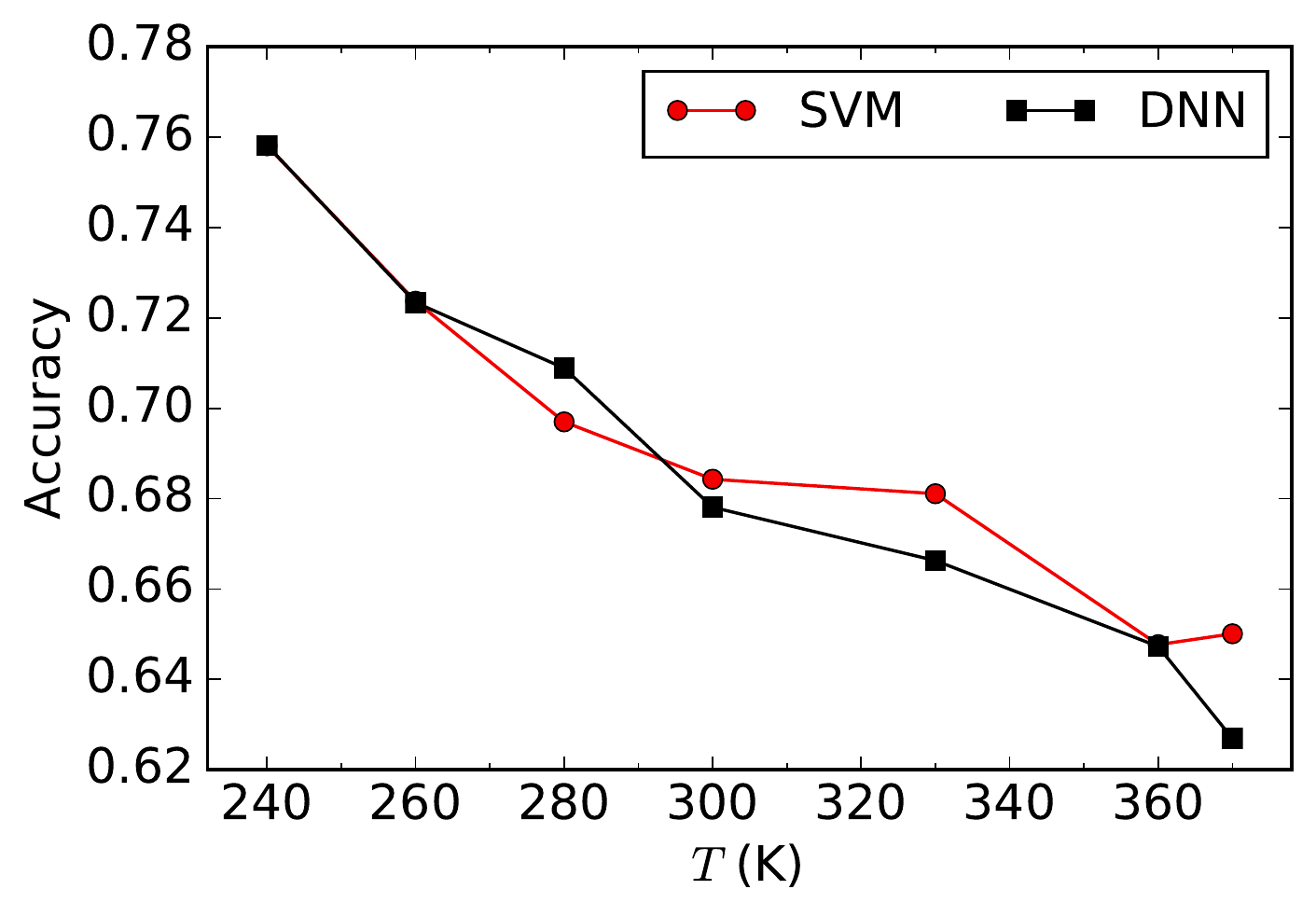}
	\vspace{-0.4cm}
	\caption{Classification accuracy of finite temperature structures for features A, training independently at each temperature a Support Vector Machine and a Deep Neural Network. The accuracy decreases with temperature, suggesting that thermal fluctuations are hindering the identification of the molecular environments.}\label{fig:errors_FTfixedT}
	\label{fig:FTfixedTerr}
\end{figure}
In order to verify this hypothesis, we performed training and classification independently at each temperature. As above, we balance the data set to prevent heavily biased solutions, keeping all LD data points and just as many HD points, randomly discarding the rest. We do not aggregate now the data from all temperatures, so in order to keep our training set large enough we reduced the number of test points use $N_{\text{test}} = \max(8000, N_{\text{left}}/4)$ for classification testing and all the remaining for data points are used for training. Indeed, in Fig. \ref{fig:FTfixedTerr} we see that the classification accuracy decreases systematically with the system temperature, indicating that the local environment described by $\vec x_i$ becomes less informative about the HD/LD character of its IS as $T$ increases. We note that since each data set size is now significantly smaller, in order to avoid an excessively complex model we reduced the number of hidden layers in the DNN from four to two, with the first one containing 80 neurons and the second one containing 30 neurons.

\subsection{The effects of short time fluctuations}\label{ssec:TimeFluctuations}
So far our analysis has been instantaneous: feature vectors and target values were generated based on geometrical properties of the environment of each molecule in a single time point. 
\begin{figure}[b]
	\includegraphics[width=0.9\linewidth]{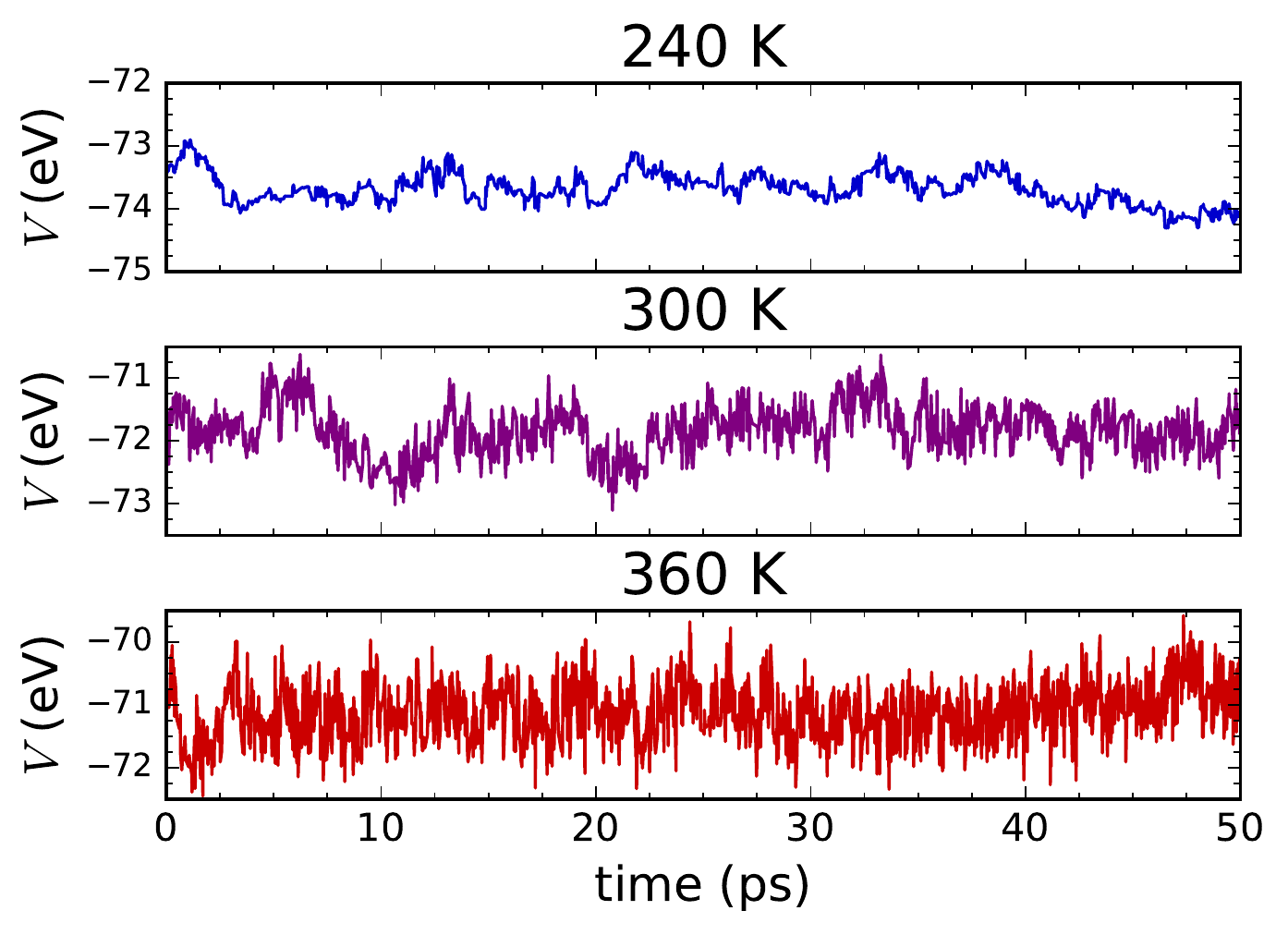}
	\vspace{-0.4cm}
    \caption{Inherent structure potential energies as a function of the simulation time for three different temperatures: $240\text{ K}$ (top), $300\text{ K}$ (center), $370\text{ K}$ (bottom). The amplitude of the short timescale ($ \lesssim 0.2$ ps) fluctuations increases with temperature.} \label{IS_EpotVStime}
\end{figure}
In Fig. \ref{IS_EpotVStime} we observe that the amplitude of the short time fluctuations of the inherent structure potential energies increase with temperature. 
Likewise, the inherent LSI as well as the components of the feature vectors fluctuate very significantly in timescales $\lesssim 500$ fs. It is therefore necessary to understand whether these fluctuations are hindering the classification of the molecular geometries. First, we observe that the FT LSIs of individual molecules are correlated with their corresponding inherent LSIs. An example is shown in Fig. \ref{fig:LSIvsTime_example} but this property holds for every molecule (see Appendix \ref{app:Fluctuations} for more examples.)
\begin{figure*}
	\centering
	\includegraphics[width=1.0\linewidth]{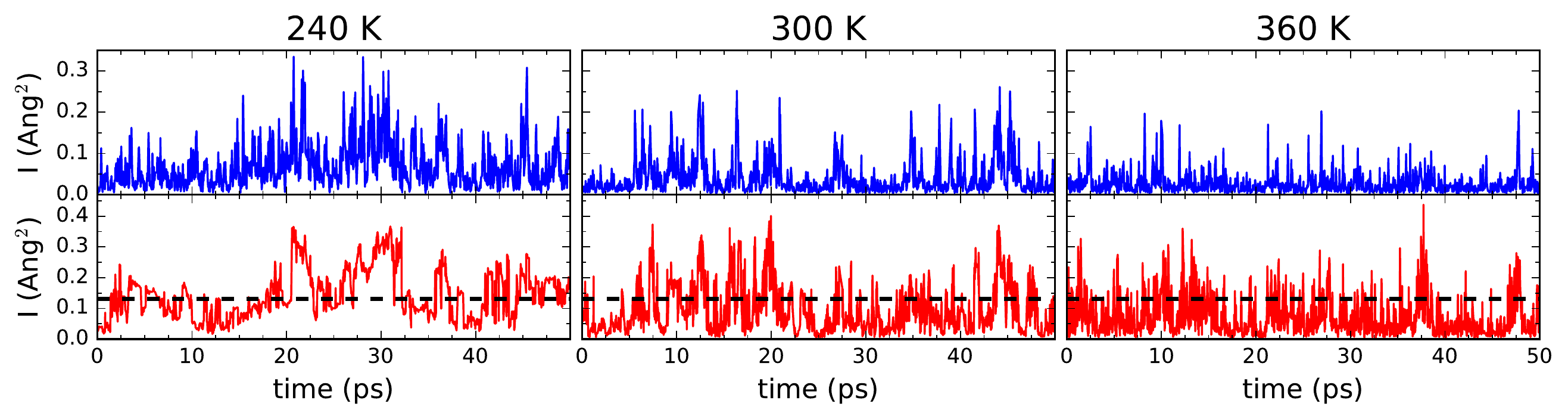}
	\vspace{-0.4cm}
	\caption{Finite temperature LSI (top) and inherent LSI (bottom) trajectories of 3 example molecules from MD simulations at different temperatures: $240$ K (left), $300$ K (center), $360$ K (right). The dashed lines show the threshold inherent LSI value $I_0=0.13$ Ang$^2$. The trajectories of these two quantities are evidently correlated. This effect is more evident at low temperatures where the inherent LSI fluctuations have smaller amplitude.}
	\label{fig:LSIvsTime_example}
\end{figure*}
So there is clearly a correspondence between FT structures and inherent structures that the ML classifier, when applied to instantaneous local data, is not able to reveal.

\subsubsection{Short-time averaging}
In order to suppress the noise introduced by the fast fluctuations and capture the time correlations we perform time averages over a short time window of length $\Delta t$ of the molecular features, which we then include in the feature vector $\vec x$. In addition we also time average the inherent LSI and generate new target values based on these averages. It would be unphysical to take time averages on a scale much larger than the intermolecular vibrations as they are a main driver for structural transitions. In addition we note that the correlation function of the fluctuation of number of H-bonds that any given molecule forms has been reported to show a characteristic fast decay time of $70-80$ fs followed by a slower decay in the $0.8-0.9$ ps range corresponding to H-bond formation and rupture \cite{Schmidt2007}. With this in mind, we time-average over a $\Delta t$ of $100$ fs and $250$ fs, values slightly above the periods of the librational modes of the water molecule and the H-bond bending mode respectively. 
\begin{figure}
	\includegraphics[width=0.9\linewidth]{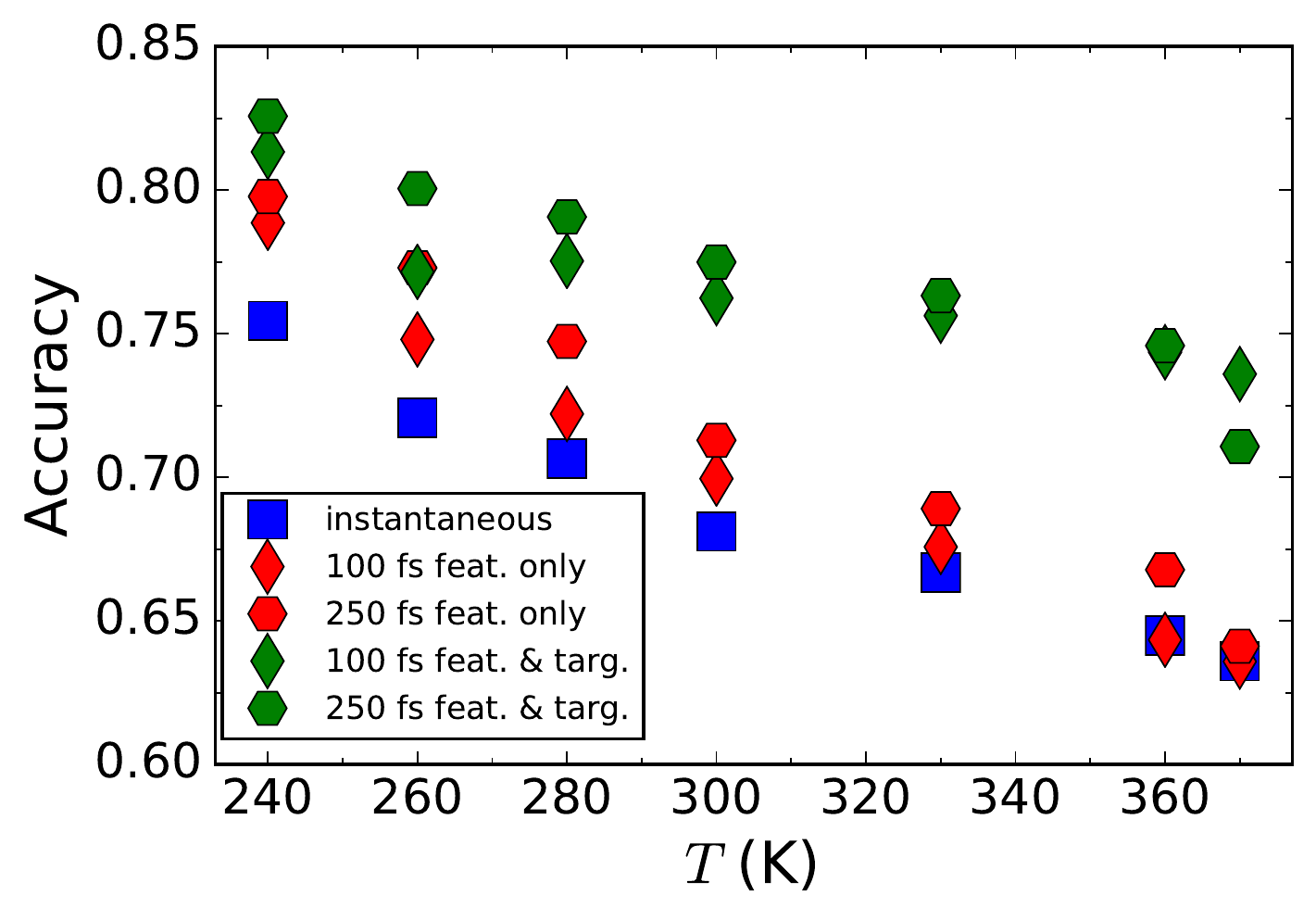}
	\vspace{-0.4cm}
	\caption{Comparison of classification accuracies using instantaneous features and targets (blue), time-averaged features and instantaneous targets (red) and time-averaged features and targets (green) obtained with a Deep Neural Network. Diamonds and hexagons represent time averages with $\Delta t = 100$ fs and $\Delta t = 250$ fs respectively.}
	\label{fig:AccVsT_TimeAveraged}
\end{figure}

We observe that training a DNN with time-averaged features included in $\vec x$ increases the classification accuracy at low temperatures by about $5\%$ where the inherent LSI fluctuates across the threshold $I_0$ at longer timescales, while at high temperatures the increase in accuracy is minimal or not significant. 
However, when the target values are constructed based on the time-averaged inherent LSI, the classification accuracy increases at all temperatures, with improvements ranging between $7\%$ and $10\%$. 
This demonstrates that the short timescale fluctuations are a fundamental reason to the difficulty of finding a one-to-one correspondence between local FT structures and inherent structures. 
It is also possible that time averaging over longer time windows would yield higher classification accuracy, but it would become questionable whether one would be finding a local (in space and in time) order parameter anymore as the time average would be taken at the vicinity of the H-bond lifetimes.

\subsubsection{Free energy and equilibrium two-state model}
With the evidences accumulated above it is worth asking the following question: can the transitions between the LD and HD inherent structures be understood in thermodynamic grounds? Let us regard the inherent LSI as an order parameter or a reaction coordinate and compute the free energy associated to the inherent structures along this coordinate. We call this quantity \textit{inherent free energy} and it is given by
\begin{equation}\label{eq:FreeEnergyFromProb}
	F(I)  = -k_B T \ln P(I)
\end{equation}
where $I$ is the inherent LSI and $T$ is the temperature of the original simulation. This is the free energy one would obtain from the partition function Eq. (\ref{eq:SWpartitionfunction}) with integrand equal to $1$.
\begin{figure}
	\centering
	\includegraphics[width=0.9\linewidth, keepaspectratio]{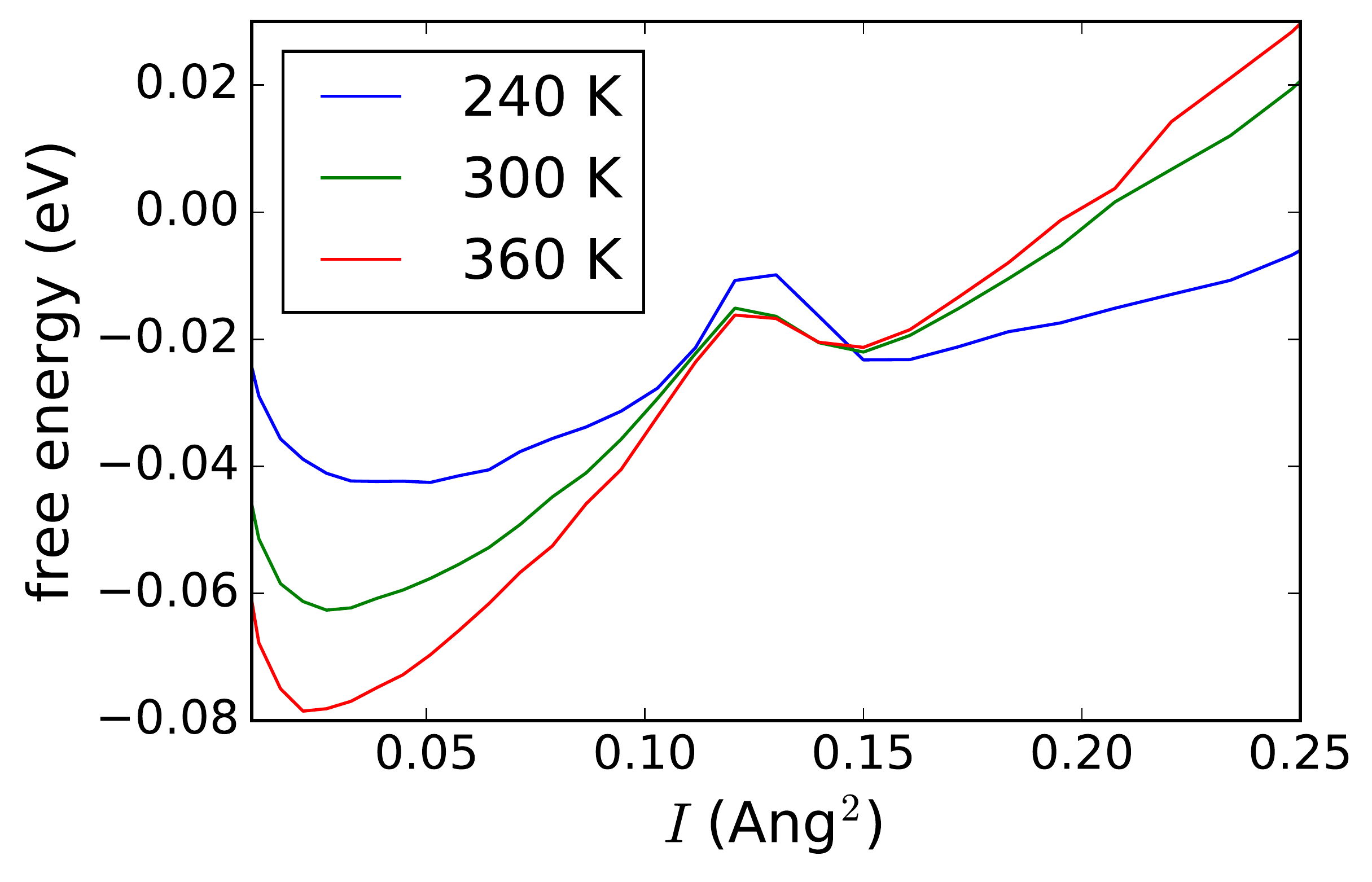}
	\vspace{-0.4cm}
	\caption{Inherent free energy surface as a function of the LSI for three different temperatures. The curves are obtained by applying Eq. (\ref{eq:FreeEnergyFromProb}) to the inherent LSI histograms, setting $T$ to the corresponding simulation temperature.}
	\label{fig:IS_FreeEnergyVsLSI}
\end{figure}
In Fig. \ref{fig:IS_FreeEnergyVsLSI} we show the inherent free energy at three different temperatures as obtained from the LSI distributions at the relaxed structures. We observe that the minima of the LD LSI minimum goes down while the HD LSI minimum goes up as $T$ increases, clearly causing a decrease of the LD population in favor of the HD-type molecules.

To go beyond this qualitative statement, the equilibrium conditions of the inherent structures can be obtained using a simple model for state transitions. Denoting the population of LD and HD molecules with $n_{\text{L}} = N_{\text{LD}}/N$ and  $n_{\text{H}} = N_{\text{HD}}/N$ respectively and the transition rates as $r_{\text{L} \rightarrow \text{H}} = r_{\text{LH}} $ and $r_{\text{H} \rightarrow \text{L}} = r_{\text{HL}}$, the rate equations of this two-state system are
\begin{align}\label{eq:RateEqs}
\frac{ dn_{\text{L}} }{dt} = r_{\text{HL}} n_{\text{H}} &- r_{\text{LH}} n_{\text{L}} \\
\frac{ dn_{\text{H}} }{dt} =  - \frac{ dn_{\text{L}} }{dt}
\end{align}
where the second equation is given by the conservation condition $n_{\text{L}} + n_{\text{H}} = 1$. Eliminating he second rate equation we obtain
\begin{equation}\label{eq:RateEqSimplified}
	\frac{ dn_{\text{L}} }{dt} = r_{\text{HL}} - \left( r_{\text{HL}} + r_{\text{LH}} \right) n_{\text{L}} .
\end{equation}
Imposing the equilibrium condition $d n_{\text{L}}/dt =0$ one obtains the equilibrium population as 
\begin{equation}\label{eq:EquilPopul}
	n_{\text{L}}^{\text{eq}} = \frac{r_{\text{HL}}}{r_{\text{HL}} + r_{\text{LH}}}.
\end{equation}
In order to make use of this equation a model for the transition rates needs to be provided. Based on collision theory \cite{Trautz1916}, we propose a phenomenological Arrhenius-type expression of the form $ r = \nu e^{ - \beta \Delta E }$ where $\Delta E$ is the activation energy (i.e. the barrier height $E^{\text{max}} - E^{\text{min}}$) and $\nu$ is the frequency of microscopic events that can lead to a state transition in the system. Using this model in Eq. (\ref{eq:EquilPopul}) and assuming that the $L \to H$ and $H \to L$ are driven by the same frequency $\nu$ we obtain
\begin{equation}\label{eq:EquilPopulFinal}
	n_{\text{L}}^{\text{eq}} = \frac{1}{1 + e^{ \beta (E^{\text{min}}_{\text{L}} - E^{\text{min}}_{\text{H}} )}}.
\end{equation}
Notice that the equilibrium population does not depend on $\nu$ or on the barrier height separating the two minima. In Table \ref{table:LD-populations} we show the difference between energy minima at these three example temperatures and we compare the populations predicted by Eq. (\ref{eq:EquilPopulFinal}), $n_{\text{FE}}$ to the populations obtained directly from the inherent LSI probability distribution, $n_{\text{LSI}}$.
\begin{table}[]
\centering
\begin{tabular}{|c|c|c|c|}
\hline
$T$ (K)                                         & 240   & 300    & 360 \\ \hline
$\Delta E_{LH}$ (meV)                & 18.54 & 39.80 & 55.98 \\ \hline
$n_{\text{LSI}}$                          & 0.335 & 0.186 & 0.159 \\ \hline
$n_{\text{FE}}$                            & 0.282 & 0.172 & 0.136 \\ \hline
$n_{\text{LSI, optimal}}$      &0.309   & 0.159 & 0.136 \\ \hline
\end{tabular}
\caption{Populations of LD component as determined by a LSI at the IS with value $< I_0 = 0.13$ Ang, obtained from imposing the equilibrium condition on the rate equations with the energy minima from the free energy curve and populations obtained from data adjusting the threshold $I_0$ to minimize the error relative to the FE estimate.}
\label{table:LD-populations}
\end{table}
We observe a qualitative agreement between the populations found from the distributions of the LSI at the IS and from the two-state model.
In addition, this model provides an alternative mechanism for determining the threshold value $I_0$: instead of using the value that minimizes $P(I)$ between the two maxima one could choose an $I_0$ that minimizes the squared-error in the populations $\frac{1}{N_T} \sum_T (n_{\text{LSI}}(T) - n_{\text{FE}}(T))^2$. This yields a value of $I=0.143$ Ang$^2$. Finally, we note that an accurate computation of the free energy would require much larger data sets than those employed here so the conclusions obtained from our free energy values are subject to error due to finite sampling size.
 
There is no clear way to obtain the frequency of events $\nu$ that regulates the transitions between the two inherent free energy minima. It is reasonable to think that the intermolecular vibrations, responsible for distortion of the H-bond network, are a main contributor to this transition.
\begin{table}[]
\centering
\begin{tabular}{|c|c|c|c|}
\hline
\multicolumn{1}{|c|}{$T$ (K)}  & \multicolumn{1}{c|}{240}  & \multicolumn{1}{c|}{300}   & \multicolumn{1}{c|}{360}   \\ \hline
$r^{-1}_{L \backslash H, \text{str}}$ (fs)                       &3270$\backslash$1270           &4190$\backslash$818                   &4980$\backslash$786 \\ \hline
$r^{-1}_{L \backslash H, \text{bend}}$ (fs)                   &882$\backslash$347               &1143$\backslash$237                   &1355$\backslash$214  \\ \hline
$r^{-1}_{L \backslash H, \text{libr}}$ (fs)                      &409$\backslash$161               &530$\backslash$110                      &629$\backslash$99  \\ \hline
\end{tabular}
\caption{L$\backslash$R inverse transition rates given by the the Arrhenius model with event frequencies of three vibrational modes in water: two slow intermolecular modes (H-bond stretch, $1.5$ THz and H-bond bend, $5.5$ THz) and one intramolecular mode (L1 libration, $11.9$ THz).}
\label{table:InverseTransitionRates}
\end{table}
Setting $\nu$ to the vibrational frequencies of in water, this Arrhenius model for the rates allows us to calculate the time between state transitions, allowing us to have an estimate of the transition timescales. The values in Table \ref{table:InverseTransitionRates} show that the time averages of Fig. \ref{fig:LSIvsTime_example} are performed over potential libration-induced transitions but not over the H-bond bend- and H-bond stretch-induced transitions.





\section{Summary and conclusions}
We proposed a method to characterize local order via supervised machine learning. The classification is based solely on the local geometrical features extracted from simulation data. Intuition about the system is necessary in order to choose feature vectors that describe appropriately the molecular environment as well as to design appropriate target values for the training data. Once the machine learning classifier is trained, it can be used determine the type of local structure associated to a given molecule and its environment, effectively providing a local order parameter. 

We demonstrate the validity of this method on inherent structures of water. First, for each data point, we choose its target value based on the instantaneous value of the inherent LSI after an appropriate thresholding. By selecting different groups of features, we observe that an order parameter can be learnt to classify inherent structures with nearly perfect accuracy provided that the intermolecular distances of the first few neighbors are included in the feature vectors. This proves that machine learning methods combined with a physically motivated feature set choice can be a very powerful tool for identification and characterization of local order in complex environments. 

We then explore the case of finite temperature structures, characterized by their corresponding inherent structures. Here classification accuracies decrease to below $75 \%$ for both of our machine learning models, challenging the prevailing view in which local order properties can be attributed to molecules based on their inherent IS \cite{Wikfeldt2011, Pettersson2014, Nilsson2015, Santra2015}. We argue that thermal effects are the main contributor to the decrease in the accuracy of our machine learning models as the molecular structures depart from the inherent structures increasingly with temperature.
There are two pieces of evidence supporting this claim. First, the classification accuracy decreases monotonically with the simulation temperature, worsening by over $11 \%$ between from lowest temperature ($240$ K) to the highest ($370$ K). And second, introducing time averages of the FT features and the inherent LSI (and hence the target values) improves the classification by approximately $10\%$. Interestingly, this improvement is more prominent at high temperatures, reducing the accuracy gap between $240$ K and $360$ K to about $7\%$.

Finally we estimate the free energy along the inherent LSI and write a phenomenological two-state equilibrium model for the inherent structures. This allows us to estimate the transition rates between IS states associated with different vibrational modes of water, providing us with sensible choices of our time average windows. 
We would like to note that it is likely that incorporating time correlations in a more sophisticated manner, such as recurrent neural networks or hidden Markov models, could further improve the identification of two classes of structures. We postpone the exploration of these ideas for future work.

In the recent years machine learning methods have excelled in image recognition, text translation, self-driving vehicles and other technological applications. They are often thought of as blackbox models with thousands of adjustable parameters that can learn features from the data but without providing understanding of the system of interest or ability to generalize outside its training data. It is primarily for these reasons that they have not become common tools of choice in the physical sciences. In this work we demonstrated that, on the contrary, machine learning can be a very insightful device only when combined with scientific understanding of the problem at hand. We expect a steep growth in the usage of these methods as scientists in these disciplines become accustomed to them. 


\section*{Acknowledgments} 
We are grateful to Philip Allen, and Jose Soler for many useful discussions. We are particularly thankful to Daniel Elton for helping set up the MD simulations and for comments on the manuscript.
A.S. and M.V.F.S. ~acknowledge support from U.S. Department of Energy grant DE-SC0001137. 
This work was partially supported by BNL LDRD 16-039 project.
This research used resources of the National Energy Research Scientific Computing Center, a DoE Office of Science User Facility supported by the Office of Science of the U.S.~Department of Energy under Contract No.~DE-AC02-05CH11231, resources of the Center for Functional Nanomaterials, which is a U.S. DOE Office of Science User Facility, at Brookhaven National Laboratory under Contract No. DE-SC0012704, as well as the Handy and LI-Red computer clusters at the Stony Brook University Institute for Advanced Computational Science.


\bibliography{BibWaterML}
\bibliographystyle{unsrt}

\appendix

\section{Support Vector Machine for binary classification} \label{app:SVM}

Support vector machines \cite{NL-SVM} have been a popular method for machine learning classification due to their algorithmic simplicity and the ease of interpretation. In this section we explain the basic formalism following \cite{Bishop-PRML}.

Let $\vec x_n = \left( x_{n,1}, \cdots, x_{n, D} \right)$ with $n=1, \cdots, N_p$ be our collection of data points in feature space and let $\tau_n \in \left\lbrace -1, +1 \right\rbrace$ be a binary label that categorizes each point that we call \textit{target value}. The set $\left\lbrace \vec x_n, \tau_n \right\rbrace_{n=1}^N$ is \textit{linearly separable} if there exists a hyperplane $y \left( \vec x \right) \equiv \vec w \cdot \vec x + b = 0$ such that $y \left( \vec x_n \right) < 0$ for points with $\tau_n = -1$ and $y \left( \vec x_n \right) > 0$ for points with $\tau_n = +1$. It is clear that a data set is, in general, not linearly separable. On the other hand, a curved surface that separates the data in the D-dimensional feature space can always be found. A convenient way of tackling the problem of finding such surface is by formulating it as a linear separation problem in a higher dimensional auxiliary space. Letting the decision boundary be now $y \left( \vec x \right) \equiv \vec w \cdot \vec \phi \left( \vec x \right) + b = 0$ where $\vec \phi \left( \vec x \right)$ is a non-linear mapping from the space of features to the auxiliary space, the \textit{margin} is defined as the normal distance from closest point to the decision surface, $\tau_n y \left( \vec x_n \right) / \lvert \lvert \vec w \rvert \rvert$. The desired surface then maximizes the margin over $\vec w$, $b$ and $n$. Taking advantage of the fact that $y \left( \vec x \right)$ is invariant under rescaling $\left(\vec w, b \right) \rightarrow \left(\alpha \vec w, \alpha b \right)$, the points closest to the surface can be set to satisfy $\tau_n y \left( \vec x_n \right) = 1$. This turns the problem of maximizing the margin into maximizing $1/ \lvert \lvert \vec w \rvert \rvert$, which is in turn equivalent to the quadratic programming problem of minimizing $\lvert \lvert \vec w \rvert \rvert ^2$ under the constraints 
\begin{equation}\label{eq:ExactClassConst}
	\tau_n y \left( \vec x_n \right) \geq 1
\end{equation} 
which ensure exact classification. Introducing the constraints via Lagrange multipliers $a_i$ we are left with the Lagrangian function
\begin{equation}\label{eq:LagrangianHardMarginSVM}
L_{\text{hard}} = \lvert \lvert \vec w \rvert \rvert ^2 - \sum_{n=1}^N a_n \left(\tau_n \left(\vec w \cdot \vec \phi \left( \vec x_n \right) \right) \right)
\end{equation} 
where the minus sign between the two terms denotes that we are minimizing with respect to $\vec w$ and $b$ and maximizing with respect to $a_n$. Defining the \textit{kernel function} as the scalar product in the auxiliary space
\begin{equation}\label{eq:DefKernel}
k \left( \vec x_1, \vec x_2 \right) = \vec \phi \left(\vec x_1 \right) \cdot  \vec \phi \left(\vec x_2 \right)
\end{equation}
and setting to zero the derivatives with respect to $\vec w$ and $b$, the Lagrangian can be rewritten in its dual representation 
\begin{equation}\label{eq:DualLagrangianHardMarginSVM}
\tilde L_{\text{hard}} = \sum_{n=1}^{N_p} a_n \left(1 - \sum_{m=1}^{N_p} a_m \tau_m \tau_m k \left( \vec x_n, \vec x_m \right) \right)
\end{equation} 
now subject to the constraints $a_n \geq 0$ and $\sum_{n=1}^{N_p} a_n \tau_n = 0$. Notice that the Lagrangian now depends on the kernel function and no longer on the explicit mapping to the auxiliary space $\phi$. In practice the user specifies a kernel function, which may have adjustable hyperparameters (this is the case of the RBF kernel described in Appendix \ref{app:RBF}). After the parameters $a_n$ have been optimized, the decision boundary can be expressed as a function of the kernel function as 
\begin{equation}\label{eq:DecBoundKernel}
	y \left( \vec x \right) = \sum_{n=1}^{N_p} a_n \tau_n k \left(\vec x_n, \vec x \right) + b
\end{equation}

This method provides a formalism to find a curved decision boundary. The decision surface, however, is at risk of being highly overfitted to the training data points and hence be very uninformative about the structure of the data distribution, achieving very limited predictive power. The idea of \textit{soft margin} relaxes the condition of strict separability, finding a hypersurface that approximately separates the data set while allowing for some misclassified points. Defining the slack variables as 
\begin{equation}\label{eq:SlackVar}
	\zeta_n=
	\begin{cases}
		0, \text{if } \vec x_n \text{ is correctly classified} \\
		\lvert y \left( \vec x_n \right) - \tau_n \rvert, \text{otherwise}
	\end{cases}
\end{equation}
the exact classification constraints (\ref{eq:ExactClassConst}) are replaced with
\begin{equation}\label{eq:SoftClassConst}
	\tau_n y \left( \vec x_n \right) \geq 1- \zeta_n
\end{equation} 
Now we can include a penalty for misclassification into the Lagrangian (\ref{eq:LagrangianHardMarginSVM}), controlled by the \textit{box parameter} C, as well as the Lagrange multipliers $\mu_i$ for the constraints $\zeta_i > 0$, resulting in
\begin{align}\label{eq:LagrangianHardMarginSVM_penalty}
	\begin{split} 
		L_\text{soft} &= \lvert \lvert \vec w \rvert \rvert ^2 + C \sum_{n=1}^{N_p} \zeta_n \\ 
			&- \sum_{n=1}^{N_p} a_n \left(\tau_n y \left( \vec x_n \right) -1 + \zeta_n \right) - \sum_{n=1}^{N_p}  \mu_n \zeta_n.
	\end{split}
\end{align} 
It can be shown that the soft margin Lagrangian can also adopt a dual representation with the same functional form as (\ref{eq:DualLagrangianHardMarginSVM}) but subject to different constraints, namely $ 0 \le a_n \le C$ and $\sum_{n=1}^{N_p} a_n \tau_n =0$. Now the Lagrange multipliers $a_n$ need to be bounded below the user-specified box parameter.

\subsection{Gaussian kernel}\label{app:RBF}
One essential piece of the SVM is the kernel function. When a non-linear kernel function is chosen, the decision boundary, calculated via Eq. (\ref{eq:DecBoundKernel}), is a curved surface in D-dimensional space. Gaussian kernels of the form
\begin{equation}\label{eq:RBF}
	K(\vec{x}, \vec{x}') = \exp \left( -\frac{\|\vec{x} - \vec{x}'\|^2}{2\sigma^2} \right)
\end{equation}
are commonly used in a variety of machine learning algorithms. This function contains a free (hyper)parameter: the Gaussian width $\sigma$. If $\sigma$ is small, the Gaussian decays rapidly as $\vec x$ moves away from $\vec x'$. On the other hand, when $\sigma$ is large, $K$ varies slowly as $\vec x$ moves away from $\vec x'$. 
For this reason, small $\sigma$ values can adjust the surface better to the training data though at the risk of overfitting and losing predictive power. A very large $\sigma$, however, could make it difficult for the training algorithm to properly fit the training data producing larger errors and, in the worst case, convergence problems. For these reasons, tuning $\sigma$ properly is a crucial step during while training this model.

\section{Neural Networks for classification}\label{app:DNN}
Here we introduce the concepts of neuron and neural network, show how they can be used for classification and explain how the network is trained. We follow mostly the notation in \cite{McKay-ITLA}, where we point the interested reader for an in-depth discussion.

A neuron is a unit that provides an output $y$ for any given input vector $\vec x$. It contains weights $w_1, \cdots , w_D$ associated to each input dimension and a bias parameter $b$. 
\begin{figure}
	\begin{center}
		\includegraphics[width=0.8\linewidth, keepaspectratio]{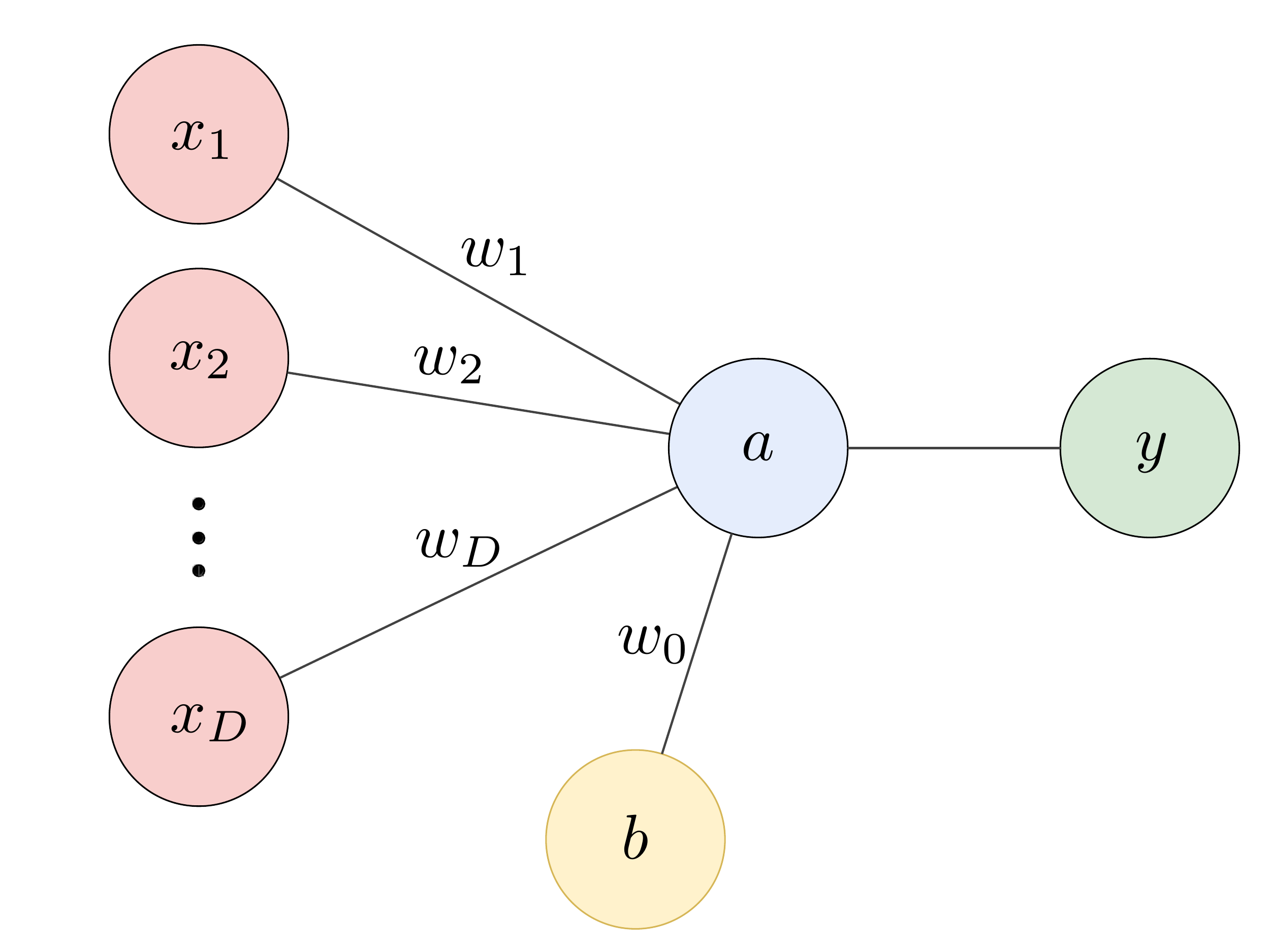}
		\caption{Graphical representation of one neuron. Activation $a$ is obtained from the features $\vec x$ by a linear mapping. The activity is the output of the neuron, given by the nonlinear function $y(a)$.}
		\label{fig:OneNeuron}
	\end{center}
\end{figure}	
The \textit{activation} of the neuron given an input $\vec x$ is a linear function given by 
\begin{equation}\label{eq:activation}
 a = \sum_i w_i \: x_i + b.
\end{equation}
So the neuron can be seen as a linear function, which could in turn be used to perform linear binary classification if we regard the sign of $a$ as to identify each class of data points upon an appropriate selection of the weights and the bias parameters. In addition, this can be generalized by introducing a (generally non-linear) \textit{activity} function $y \left( a \right)$. By choosing suitable activity functions such as $f(a) = 1/( 1 + e^{-a} )$ or $f(a) = \tanh(a)$ one can mimic the biological behavior of a neuron, which propagates information by enabling an action potential whose value is bounded above and below. Interestingly, the virtue of the abstract neuron presented here relies on two ideas: the use of nonlinear activities and combination of multiple neurons together in a feed-forward network topology (see Fig. \ref{fig:NNdiagram}). 
Let us imagine a collection of $l_1$ neurons, each acting on the input data  $\vec x$ of dimension $D$ and with output 
\begin{equation}\label{eq:DNNlayer1}
 \vec z^{(1)}  ( \vec x )  = f(w^{(1)} \vec x + \vec b^{(1)})
\end{equation}
where $w^{(1)}$ is a $D \times l_1$ matrix with rows containing the weights corresponding to each neuron and $\vec b^{(1)}$ is a $1 \times l_1$ vector containing the biases of all neurons. Now each output unit $ \vec z^{(1)}_i$  is a nonlinear function of the input data $\vec x$. These output units can themselves be input to a new set of $l_2$ neurons, producing the output 
\begin{equation}\label{eq:DNNlayer2}
 \vec z^{(2)}  ( \vec z^{(1)} ) = f(w^{(2)} \vec z^{(1)} + \vec b^{(2)})
\end{equation}
By adding this second layer of neurons we obtain a nonlinear function acting on a nonlinear function of the input data $\vec x$. More layers can be added to systematically create a more complex nonlinear function of the input $\vec x$. Proceeding with $H$ hidden layers, $z^{(h_k)}$, and terminating this sequence with an output layer with linear activity, we obtain
\begin{equation}\label{eq:DNNlayerOut}
\vec y ( \vec z^{(H)} ) = w^{(\text{out})}  \vec z^{(H)} +  \vec b^{\text{(out)}}.
\end{equation}
This formalism can tackle classification problems of great complexity with nonlinearly separable data if the weights and biases are properly adjusted. The number of such adjustable parameters in the network is given by the sum of the number of weight parameters and bias parameters in the connections between adjacent layers. Denoting the number of hidden layers by $H$ and the number of units in each layer by $l_k$, we have
\begin{equation}\label{eq:NN_Nparams}
	N_{\text{p}} = \sum_{k=0}^{H} (l_{h_k} + 1) l_{h_{k+1}} 
\end{equation}
where $k=0$ corresponds to the input layer and $k=H+1$ to the output layer.
Notice that the output $\vec y$ is a composition of functions
\begin{equation}\label{eq:DNNfunction}
\vec y  \left( \vec x \right) =  \cdots  z^{(3)} ( z^{(2)} ( z^{(1)} ( \vec x ) ) )
\end{equation}
and therefore the output can be written as a function of $\vec x$ with a parametric dependence on the biases and weights of each layer.

Let us briefly discuss how the free parameters of the network are optimized. To simplify the discussion, we assume in the following that the output layer consists of a single unit $y$ with values between $0$ and $1$ and that the target values for binary classification are $t_n =0,1$. \footnote{The extension to multiclass classification is straightforward by including as many output units as classes and interpreting the output value of each unit as the probability that the data point belongs to that class.} We define the error function
\begin{equation}\label{eq:CrossEntropy} 
	E = - \sum_{n=1}^{N_p} \left[ t_n \ln y \left( \vec x_n \right) + (1-t_n) \ln (1 - y \left( \vec x_n \right) ) \right]
\end{equation}
which is the relative entropy between the probability distribution generated by the network, $(y, 1-y)$ and the ``observed" probability distribution of the data $(t, 1-t)$. This is a function of all weights and biases in the network and it is bounded by zero from below, corresponding to perfectly matching network prediction with the observed data. This error function can be minimized by ``descending" it along the direction of the gradient. This requires computing derivatives of this function with respect to the network parameters. Luckily, since the output $\vec y$ is a composition of analytically known functions
\begin{equation}\label{eq:DNNfunctionComposition}
y  \left( \vec x ; \lbrace w, b \rbrace \right) =  \cdots  z^{(2)} ( z^{(1)} ( \vec x; \lbrace w^{(1)}, b^{(1)} \rbrace ); \lbrace w^{(2)}, b^{(2)} \rbrace )
\end{equation}
derivatives can be readily calculated applying the chain rule. This gradient can computed for each data point $n$, allowing to adjust the weights and the biases a small amount in the direction of descending gradient for $\vec x_n$. Alternatively, in order to make this computation more efficient, instead of computing the gradient at each data point one can calculate these gradients for batches of data points, averaging. This procedure is called batch gradient descent, which we used in this work.

\section{Time fluctuations of the LSI}\label{app:Fluctuations}

We argued above that short time fluctuations are responsible (at least partly) of the difficulty of identifying an instantaneous FT structure with its corresponding low-LSI or high-LSI inherent structure, even if using a very flexible pattern recognition device like a DNN. 

\begin{figure*}
	\centering
	\includegraphics[width=0.9\linewidth]{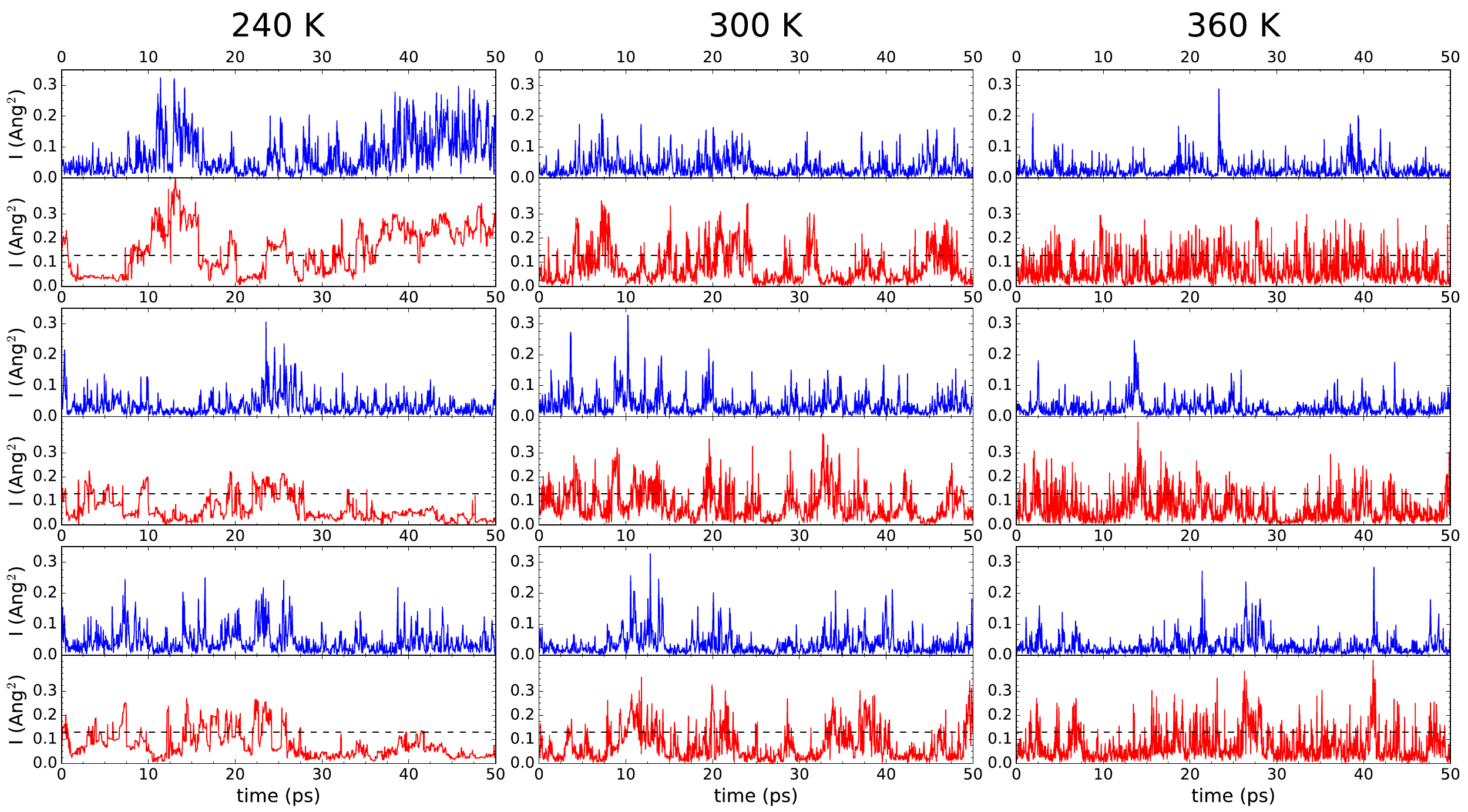}
	\caption{LSI trajectories at FT (top, blue), LSI at the IS (bottom, red) for 9 example molecules at 3 different temperatures: $240$ K (left),  $300$ K (center),  $360$ K (right). The dashed lines show the threshold LSI value  $I_0=0.13$ Ang$^2$.}\label{fig:LSIvsTime_3molecules}
\end{figure*}

\begin{figure*}
	\centering
	\includegraphics[width=0.9\linewidth]{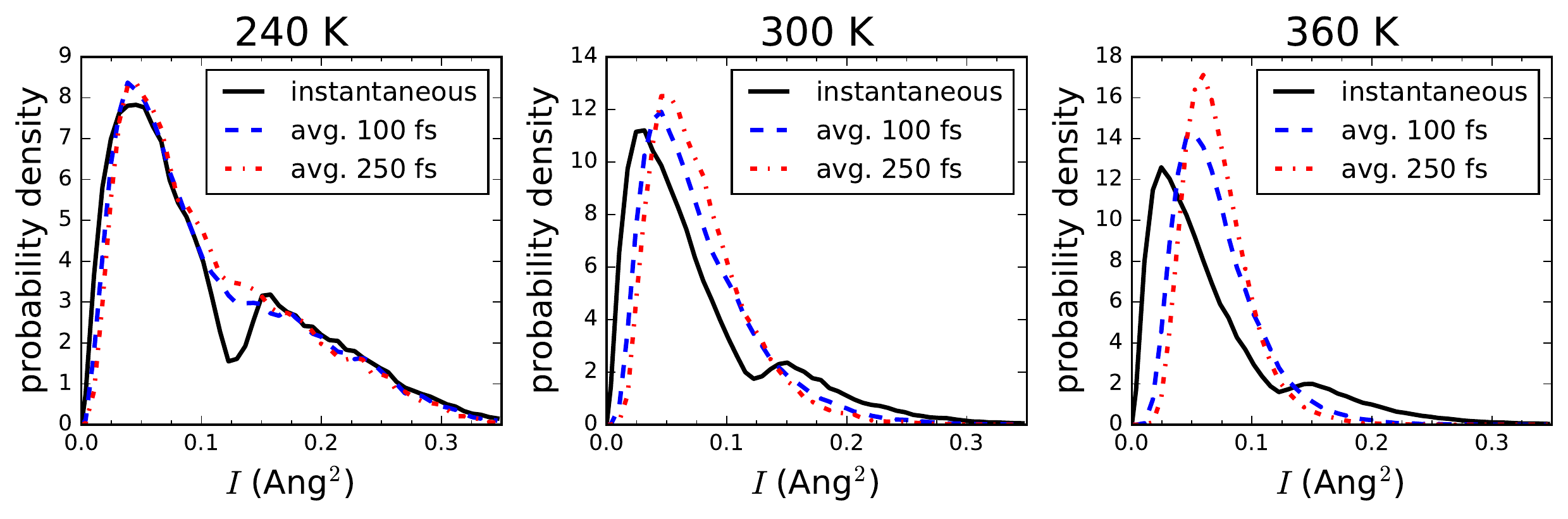}
	\caption{Histograms of instantaneous inherent LSI (solid black) and time averaged inherent LSI over $100$ fs (dashed blue) and $250$ fs (dashed-dotted red). The bimodal character of the probability distribution is disappears upon time averaging, revealing that this bimodality is a short-time effect.}
	\label{fig:LSI_histograms_TimeAveraged}
\end{figure*}

The LSI plays a central role in our analysis since it is our basis to define the target values for supervised machine learning. To show further evidence on our claim about the correlation between the FT LSI and the inherent LSI discussed in \ref{ssec:TimeFluctuations}, in Fig. \ref{fig:LSIvsTime_3molecules} we show a LSI trajectories of 3 example molecules at 3 different temperatures.
In addition we show in Fig. \ref{fig:LSI_histograms_TimeAveraged} the histograms of the inherent LSI as well as the histograms of the time-averaged inherent LSI for our choices of $\Delta t = 100$ fs and $250$ fs. The bimodal character vanishes upon averaging, this effect being more pronounced at higher temperatures.
It is therefore no longer well defined to split the distribution into two, one corresponding to each type of molecular environment.

\end{document}